\documentclass[%
 aip,
rsi,%
 amsmath,amssymb,
 reprint,%
]{revtex4-1}

\usepackage{graphicx}
\usepackage{dcolumn}
\usepackage{bm}

\begin{document}

\preprint{AIP/123-QED}

\title[Fundamental aspects of noise in analog-hardware neural networks]{Fundamental aspects of noise in analog-hardware neural networks}

\author{N. Semenova}
\email{nadezhda.semenova@femto-st.fr}
\affiliation{D\'{e}partement d'Optique P. M. Duffieux, Institut FEMTO-ST,  Universit\'e Bourgogne-Franche-Comt\'e CNRS UMR 6174, Besan\c{c}on, France.}%
\affiliation{Department of Physics, Saratov State University, Astrakhanskaya str. 83, 410012 Saratov, Russia.}%

\author{X. Porte}
\affiliation{D\'{e}partement d'Optique P. M. Duffieux, Institut FEMTO-ST,  Universit\'e Bourgogne-Franche-Comt\'e CNRS UMR 6174, Besan\c{c}on, France.}%

\author{L. Andreoli}
\affiliation{D\'{e}partement d'Optique P. M. Duffieux, Institut FEMTO-ST,  Universit\'e Bourgogne-Franche-Comt\'e CNRS UMR 6174, Besan\c{c}on, France.}%

\author{M. Jacquot}
\affiliation{D\'{e}partement d'Optique P. M. Duffieux, Institut FEMTO-ST,  Universit\'e Bourgogne-Franche-Comt\'e CNRS UMR 6174, Besan\c{c}on, France.}%

\author{L. Larger}
\affiliation{D\'{e}partement d'Optique P. M. Duffieux, Institut FEMTO-ST,  Universit\'e Bourgogne-Franche-Comt\'e CNRS UMR 6174, Besan\c{c}on, France.}%

\author{D. Brunner}
\affiliation{D\'{e}partement d'Optique P. M. Duffieux, Institut FEMTO-ST,  Universit\'e Bourgogne-Franche-Comt\'e CNRS UMR 6174, Besan\c{c}on, France.}%

\date{\today}

\begin{abstract}

We study and analyze the fundamental aspects of noise propagation in recurrent as well as deep, multi-layer networks. 
The main focus of our study are neural networks in analogue hardware, yet the methodology provides insight for networks in general. 
The system under study consists of noisy linear nodes, and we investigate the signal-to-noise ratio at the network's outputs which is the upper limit to such a system's computing accuracy.
We consider additive and multiplicative noise which can be purely local as well as correlated across populations of neurons.
This covers the chief internal-perturbations of hardware networks and noise amplitudes were obtained from a physically implemented recurrent neural network and therefore correspond to a real-world system.
Analytic solutions agree exceptionally well with numerical data, enabling clear identification of the most critical components and aspects for noise management. 
Focusing on linear nodes isolates the impact of network connections and allows us to derive strategies for mitigating noise. 
Our work is the starting point in addressing this aspect of analogue neural networks, and our results identify notoriously sensitive points while simultaneously highlighting the robustness of such computational systems.

\end{abstract}

\maketitle

\begin{quotation}

The implementation of neural networks in classical, digital hardware has been identified as a serious performance bottleneck. 
This strongly boosts efforts to realize neural networks in analogue systems hosting the physical links between neurons. 
Such systems promise to significantly improve speed and energy efficiency, yet they are fundamentally prone to noise originating from their analogue components. 
Here, we study for the first time how such noise propagates through recurrent and deep, multi-layer networks, and derive an analytical description for such systems. 
While noise certainly cannot be fully suppressed, our work shows that analogue neural networks can be surprisingly robust. 
From an architecture point of view, it turns out that the system's sensitivity to noise is mainly located in the first and final layers. 
Surprisingly, only noise correlated across populations of neurons proofs to have crucial effects for information propagation through the network; meanwhile purely local, uncorrelated noise can mostly be ignored or mediated. 
These are indispensable insights for designing high future analogue hardware neural networks.

\end{quotation}

\section{\label{sec:Intro}Introduction}

Networks are the underlying principle for countless physical systems and information processing concepts. 
Particularly in computing, they support a wide range of powerful algorithms such as Hopfield \cite{Hopfield1982} and neural \cite{McCulloch1943} networks as well as in Ising machines \cite{Utsunomiya2011}. 
Especially neural networks have recently resulted in a revolution of modern computing \cite{LeCun2015}. 
By now, deep neural networks achieve super-human performance in computational tasks previously deemed un-solvable by computers. 
They are now established as indispensable and as the current disruptive computing technology.
 
A fundamental aspect of neural networks is the propagation of information between nodes along weighted connections, making parallelism essential in neural network computing concepts. 
However, our digital computing architectures are serial and spatially separate memory from the location of information transformation. 
As each of a neural network's connection-weight corresponds to one value of memory, the ratio between floating point operations (FLOP) and memory access (byte) is fundamentally skewed when compared to classical computing. 
As a consequence, high-performance neural network computing is today performed on special-purpose integrated circuits (SAICs) such a graphic processors (GPU) or Google's tensor processing unit (TPU) \cite{Jouppi2017}.

Maximal computing performance can therefore only be achieved if neural networks are fully hardware implemented. 
In such systems, the overhead due to serial communication is avoided: each neuron corresponds to a simple nonlinear component, each connection to a direct physical link. 
Motivated by the potential benefits, research activity along these lines has lately exploded. 
Novel physical components such as lasers \cite{Brunner2013a}, memristors \cite{Tuma2016} and spin-torque oscillators \cite{Torrejon2017} have been shown to serve as excellent analogue neurons. 
Simultaneously, mostly optical concepts for parallel networks based on holography \cite{Psaltis1990}, diffraction \cite{Bueno2018,Lin2018}, integrated networks of Mach-Zehnder modulators \cite{Shen2016b} and wavelength division multiplexing \cite{Tait2017} have been demonstrated. 
Several review articles summarize various trends inside this area \cite{VanderSande2017,Tanaka2019,Schuman2017,Hasler2013}.

It is equally clear that, besides the potentially large benefits, such parallel and analogue hardware platforms face new, fundamental challenges. 
Among the most basic differences to digital circuits, is that the noise of individual components propagates along the network. 
Consequently, an important concern is that such systems might ultimately succumb to the detrimental impact of noise, rendering computing impossible. 
Regardless of its relevance, no previous study analyses this aspect in detail, but focus on noise propagation in serial analogue circuits \cite{Sarpeshkar1998} and the interaction between noise and learning \cite{Murray1991}. 
Here, we study fundamental properties of noise and propagation along connections of networks, considering feed forward neural networks (FNNs) as well as recurrent neural networks (RNNs). 
We consider correlated and uncorrelated, multiplicative and additive noise present in individual neurons and use noise amplitudes extracted from a physical experiment \cite{Bueno2018}. 
This covers a large range of noise sources possibly encountered in analogue hardware neural networks. 
We assume linear neurons in order to exclusively focus on the mixing of noise due to network-connections. 
Analytical dependencies of noise propagating are derived, providing a clear understanding of the relevant processes. 
Finally, we provide guidelines for hardware architectures and identify the critical points where such systems are most vulnerable.
 
\section{\label{sec:ANN}Analogue hardware neural networks}

Neural networks can be found in multiple configurations which can be categorized by their connectivity.
 In general, a neuron's internal state $x$ sums input from other neurons according to connectivity weights $\mathbf{W}$.
 The internal state is then mapped onto its output $y'$.
 Figure \ref{fig:one_node}(a) schematically illustrates such a unit, also referred to as a perceptron. 
 For neuron $i$, the set of governing equations is
\begin{eqnarray}
x_i =& \sum_{j}^{I} W_{i,j} x_j + b_i, \\ \label{eq:simplNeuron}
y'_i =& f\left( x_i \right). \label{eq:NodeSelectB}
\end{eqnarray}
\noindent In this simple description the number of neurons is $I$, and $b_{i}$ is a constant bias offset.
 As previously mentioned, we will restrict this work to a linear mapping according to $f(x)=\alpha x$.
 Importantly, if the neuron is located in the first layer, then its input $x_j$ is replaced by th system's input signal $u$.

This set of equations forms the basis of our description.
 It will be adopted to different architectures, while neurons will include sources of noise.
 Typically, we normalize connection matrices to their largest eigenvalue, which (i) maintains the signal amplitude for FNNs, or (ii) makes RNNs more comparable to FNNs.
 In FNNs, connections $\mathbf{W}$ are established in a cascaded manner, connecting the neurons of a layer to the neurons of a previous layers.
 In RNNs, connections $\mathbf{W}$ of a hidden layer establish connections between neurons from the same layer.
 This introduces a temporal context in the RNNs state, and such systems feature short term memory and can be employed to process temporal information \cite{Jaeger2002ShortNetworks}.

In addition, we restrict our treatment to connection matrices with purely positive connections.
 This was (i) motivated by our hardware-system, and (ii) corresponds to a restriction commonly found in optical networks \cite{Psaltis1990}.
 Finally we only utilize single neurons in the first (input) and final (output) layer.

\subsection{\label{sec:ANN_noise}Different sources of noise}

\begin{figure}[t]
\center{\includegraphics[width=1\linewidth]{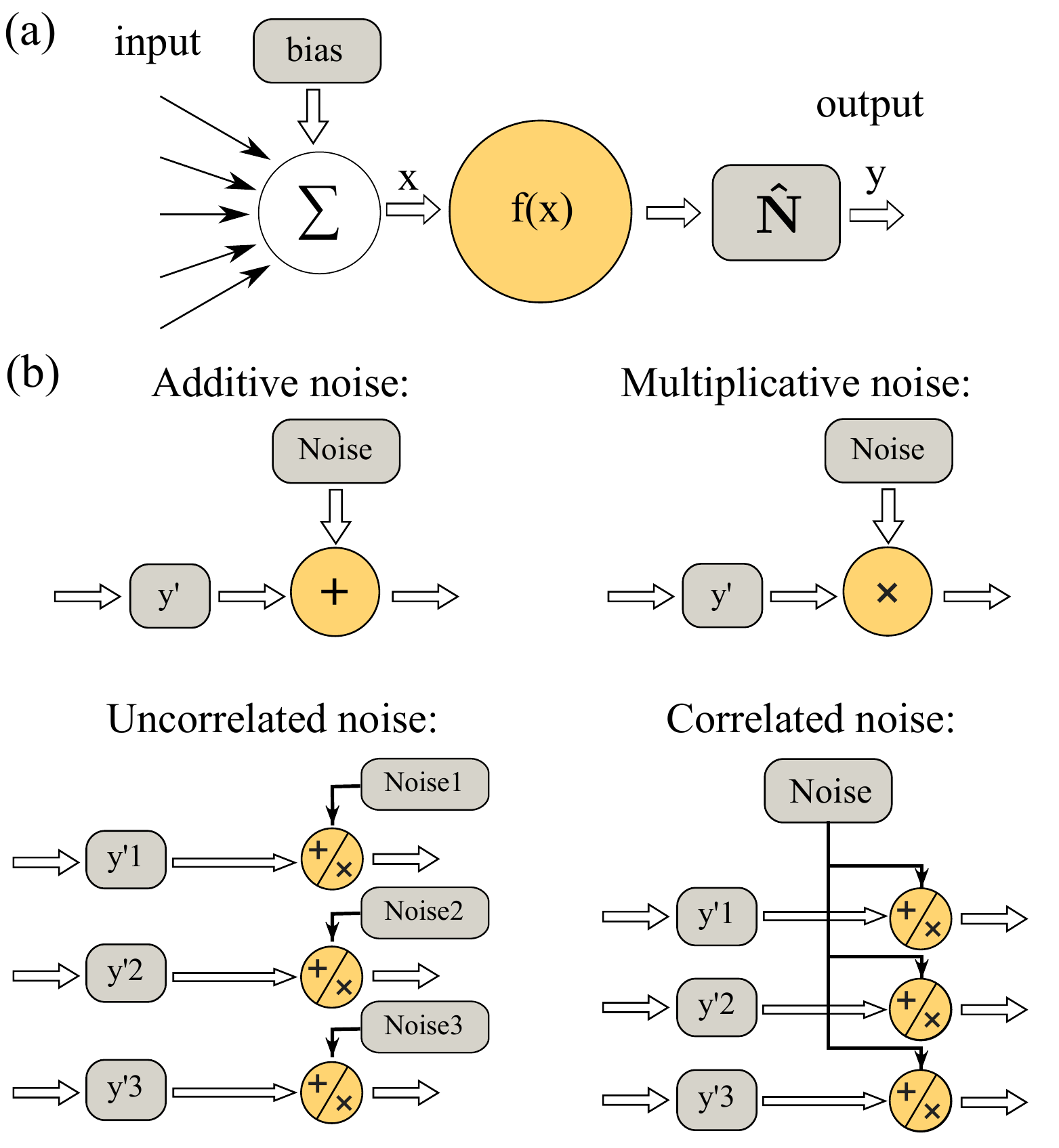}}
\caption[]{Schematic diagram showing the signal propagation through one noisy neuron.}
\label{fig:one_node}
\end{figure}

For simplicity we start by considering sources and the impact of noise upon a single neuron.
 A schematic illustration of the relevant processes are shown in Fig.~\ref{fig:one_node}(a). 
 After a neuron's internal state $x$ has been mapped onto its noiseless output $y'$ according to Eq. (\ref{eq:NodeSelectB}), the signal becomes perturbed by noise operator $\mathbf{\hat{N}}$:
\begin{equation}\label{eq:one_neuron}
y=y'+\mathbf{\hat{N}}(y').
\end{equation}
\noindent  We focus on two noise families, see Fig. \ref{fig:one_node}(b).
 For additive noise, operator $\mathbf{\hat{N}}$ does not depend on incoming signal $y'$, i.e. has properties and characteristics independent of the neuron's internal state. 
 If, however, noise is multiplicative, then operator $\mathbf{\hat{N}}$ depends on $y'$ and hence on $x$.

In this work we consider only white Gaussian noise sources according to:
\begin{equation}
\begin{array}{l}
\text{additive noise: } \mathbf{\hat{N}}(y')=\sqrt{2D_A}\xi_A \nonumber\\
\text{multiplicative noise: } \mathbf{\hat{N}}(y')=y'\cdot\sqrt{2D_M}\xi_M, 
\end{array}
\end{equation}
where $\xi_A$ and $\xi_M$ are sources of white Gaussian noise with zero expectation value and a variances of 1.
 The variance of the noise operator is controlled by noise intensities $D_A$ and $D_M$. 

In our description, each $\xi$ value can depend on four indexes: (i) - time $t\in[1,T]$ ($T$ is the length of the input sequence), (ii) - repetition number $k\in[1,K]$ (needed for SNR calculation), (iii) - layer number $n\in [1, N]$ and (iv) - a neuron's index $i\in [1,I_n]$ in a layer $n$.
 At this stage we need to contemplate about the potential impact typical hardware circuitry can impose.
 Most fundamentally speaking, component (neuron) internal processes will be locally unique and hence result in noise which is different for each neuron at each instant in time.
 The describing term depends on each index $t,k,n,i$ and hence we refer to this type as \textit{uncorrelated noise}.
 A schematic circuit noise-model for uncorrelated noise is illustrated in Fig. \ref{fig:one_node}(b).
 In the present work we denote uncorrelated noise by the letter $'U'$.
 Uncorrelated additive noise, for example, is governed by $\xi^U_A=\xi^U_A(t,k,n,i)$ with $D^U_A$ as its noise intensity. 

Another feature commonly encountered in hardware circuits is that a few elements impact / control the overall circuit's state.
 A simple example would be a circuit's supply voltage or its temperature.
 Noise in such central components will perturb large fractions of the circuit in a comparable manner. 
 In our considerations such noise is the same inside one layer and therefore only depends on time $t$, number of repetition $k$ and the layer's number $n$. 
 It is therefore correlated across neuron populations, and \textit{correlated noise} is denoted by the letter $'C'$.
 A schematic circuit noise-model for correlated noise is illustrated in Fig. \ref{fig:one_node}(b). 
 Correlated additive noise, for example, is governed by $\xi^C_A=\xi^C_A(t,k,n)$ with $D^C_A$ as its corresponding noise intensity.
 
Each noisy neuron can therefore exhibit four different types of noise: correlated additive $\xi^C_A$ and correlated multiplicative $\xi^C_M$, as well as uncorrelated additive $\xi^U_A$ and uncorrelated multiplicative $\xi^U_M$.
 In the presence of both types of noise the output signal emitted by a single neuron at time $t$ is
\begin{equation}\label{eq:one_neuron_noise}
\begin{array}{c}
y^t_{n,i}=y'^t_{n,i}\cdot\big( 1+\sqrt{2D^U_M}\xi^U_{M n,i}\big)\big( 1+\sqrt{2D^C_{M}}\xi^C_{Mn}\big) + \\ 
\sqrt{2D^U_A}\xi^U_{An,i}+\sqrt{2D^C_A}\xi^C_{An},
\end{array}
\end{equation}
\noindent Crucially, the introduced four noise classes were motivated by findings during the careful characterization of our experiment \cite{Bueno2018}.
 We found that the different components exhibit and induce different noise characteristics.
 The camera, for example, mostly contributed uncorrelated multiplicative noise, which we attributed to timing-jitter in the device's clock.
 The illuminating laser, on the other hand, influences all neuron-states simultaneously and results in correlated multiplicative noise.
 The same type of noise is induced by the spatial light modulator, and importantly it is of significantly larger intensity and hence dominates over the laser's contribution.
 The SLM also contributed uncorrelated, additive noise, and so does the system's readout-detector.
 We have characterized each of these sources individually and in detail.
 Noise intensities used on the following are based upon the values we obtained.

\subsection{\label{sec:ANN_SNR}Signal to noise ratio}

To quantify the corruption of a signal by noise, we employ the signal-to-noise ratio (SNR)
\begin{equation}\label{eq:SNR_general}
\text{SNR}(y^t)=\frac{E(y^t)}{\sigma(y^t)},
\end{equation}
 where $y$ is the noisy signal, $E(\cdot)$ and $\sigma(\cdot)$ are the signal's mean value and standard deviation, respectively. 

There are many other well-known method for noise analysis, for example autocorrelation, cross-correlation and consistency based analysis. 
 However in the present article, we (i) know the temporal characteristics of the noise-sources, and (ii) are more interested in signal amplitude-related effects which enable a first interpretation of results with respect to a system with nonlinear neurons.
 We therefore break the overall SNR down into its values at different amplitudes $E(y)$.
 This allows to associate system performance to potential modifications by nonlinearities.
 It furthermore illustrates the importance of exploiting a hardware system's entire dynamical amplitude range.
 
We follow that strategy and procedure for characterizing deep FNNs as well as RNNs.
 The network's input signal at time $t$ is $u^t$, leading to network output $y^t$.
 We repeat the same input sequence $u^t$ for several repetitions $k\in[ 1, K]$ and obtain the sequence of the output values $y^t_k$.
 We average the obtained sequence across all $k$s and obtain the mean-response $E(y^t)=1/K\sum^K_{k=1}y^t_k$.
 The standard deviation of sequence $\sigma(y^t_k)$ is obtained following the same strategy. 

\begin{figure}[tbp]
	\center{\includegraphics[width=1\linewidth]{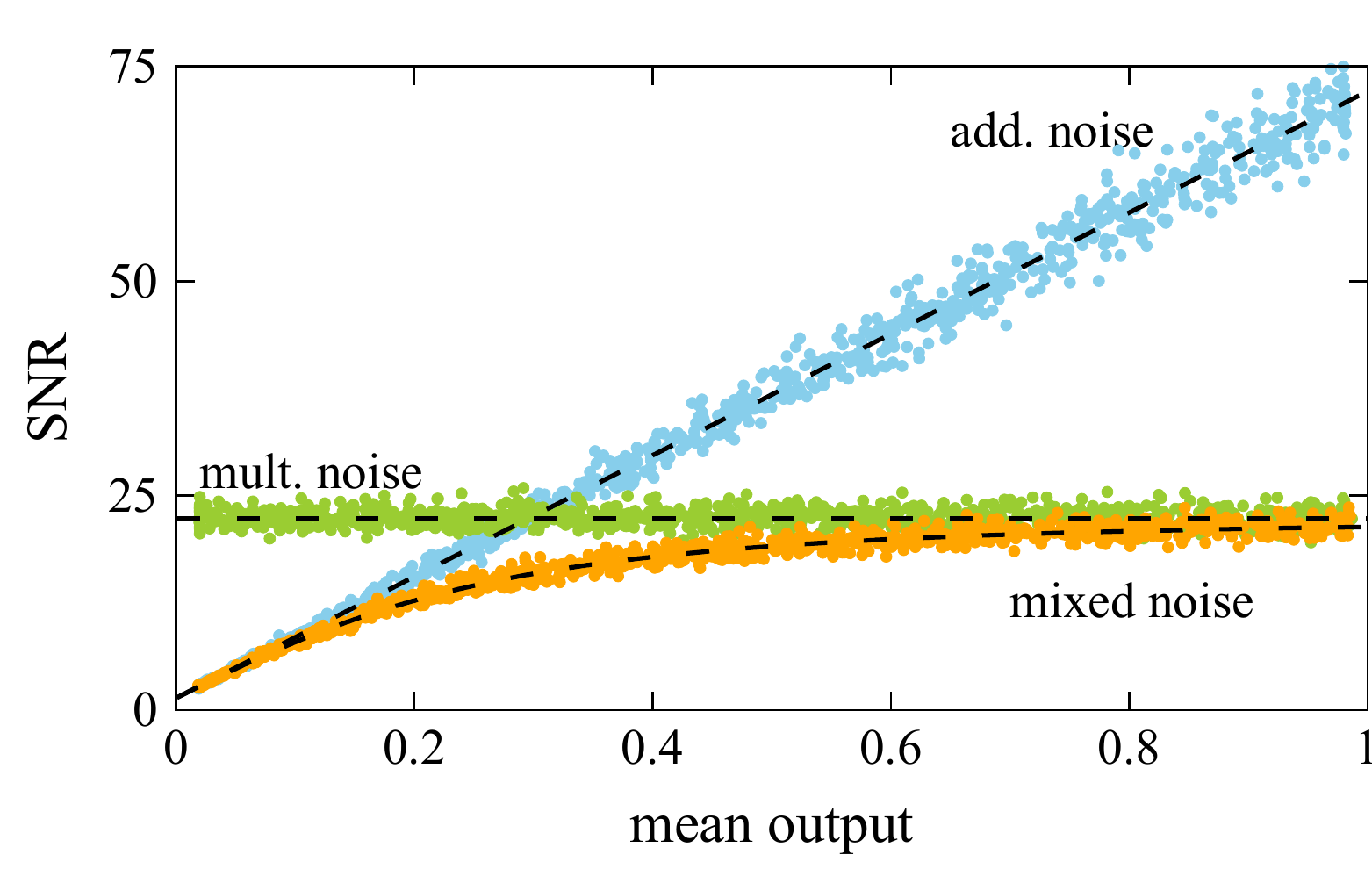}}
	\caption[]{Signal-to-noise ratio for one neuron in the case of only additive noise (blue points) with intensity $D^U_A=10^{-4}$, only multiplicative noise (green points) with intensity $D^U_M=10^{-3}$ and mixed noise (orange points) with both types of noise of the same intensities. Parameters: $\alpha=1$, $b=0.02$.}
	\label{fig:SNR_one_node}
\end{figure}

We first consider the impact of additive and multiplicative noise ($D^U_A=10^{-4}, D^U_M=10^{-3}$) on the SNR of a single neuron.
 Figure \ref{fig:SNR_one_node} shows the SNR depending on the mean output values $E(y^{t})$ for the cases of only additive (blue points), only multiplicative (geen points), as well as a combination of both noise types, hence mixed noise (orange points).

A single-neuron's SNR has features which form the basis for interpreting more complex systems.
 Due to the zero expectation value for the input signal and all noise sources, the mean value of the neuron's output is $E(y^t)=f(x)=u^t$. 
 As derived in Appendix~\ref{sec:Appendix_A}, the variance of the noisy output sequence is $\text{Var}(y^t)=\sigma^2(y^t)\approx 2D^U_A+2D^U_M f(x)^2$, leading to $\text{SNR}(y^t)=E(y^t)/\sqrt{2D^U_A+2D^U_M (E(y^t))^2}$. 
 For the case of only multiplicative noise this simplifies to $ \text{SNR}(y^t)=1/\sqrt{2D^U_M}$, which is constant for each output signal.
 As shown by the green data in Fig.~\ref{fig:SNR_one_node}, noise output signal level increase both linearly and the SNR is constant. 
 For the case of additive noise only, the SNR becomes $ \text{SNR}(y^t)=E(y^t)/\sqrt{2D^U_A}$, and the SNR of additive noise therefore increases linearly with $E(y^t)$ according to slope $1/\sqrt{2D^U_A}$, see blue data in Fig.~\ref{fig:SNR_one_node}.
 Finally, the SNR for mixed noise combines both features.
 For small output values it coincides with the SNR for additive noise, while for larger output values the SNR is limited by the constant SNR of multiplicative noise.

Before we can apply the single neuron nomenclature to networks, we need to introduce a final tool.
 Each node of an artificial neural network can exhibit some bias $b$.  
 In particular for the here considered linear system with only uni-polar (positive) connection weights, this bias can accumulate during its propagation through network layers, creating a constant output signal offset. 
 For deep FNNs, this offset increases with the number of layers.
 According to unfolding in time \cite{Williams1989}, a RNN can be mapped onto a FNN, linking the strength of a RNN's internal coupling too the depth of a corresponding FNN.
 Similar offset accumulation is therefore found when increasing a RNN's internal connection strengths.
 This offset is constant and does therefore not contribute to information processing, yet it would increase $E(y^t)$ and hence induce artefacts in $\text{SNR}(y^t)$.
 We calculate the system's response without noise on a zero-input signal $0$, and by that isolate the contribution of the constant biases.
 We refer to this value as the \textit{shifting constant} $C$, which needs to be subtracted from each output value before calculating the SNR.
 We are able to analytically describe this step for FNNs and RNNs and hence maintain the generality of our analysis.
 
 
\section{\label{sec:FNN}Noise in deep feed-forward networks}

\begin{figure}[tbp]
\center{\includegraphics[width=.8\linewidth]{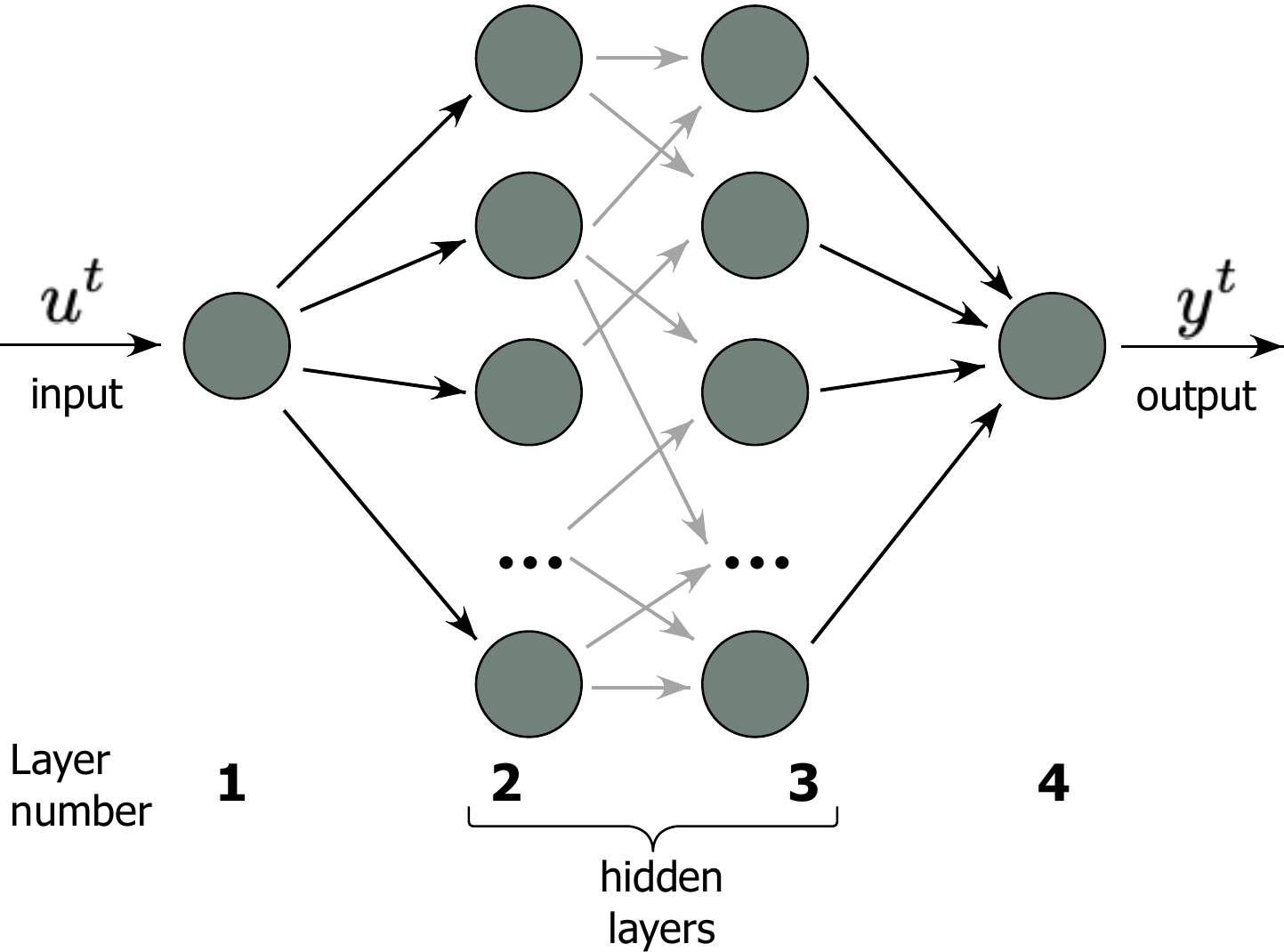}}
\caption[]{Schematic representation of a simple feed-forward neural network (FNN) consisting of $N=4$ layers.
	The central two layers are referred to as hidden layers, which each hosts $I_2=I_3=200$ neurons.
	The first and final layers consist of single neurons, $I_1=I_4=1$.}
\label{fig:FNN_scheme}
\end{figure}

We consider the FNN schematically illustrated in Fig.~\ref{fig:FNN_scheme}. 
 The network consists of four layers, with one neuron in the first and last layer $I_1=I_4=1$, and $I_2=I_3=200$ neurons in the two hidden layers. The neuron in the first layer is connected to the system's input $u^t$. 

 The system's evolution is governed by
\begin{eqnarray}
x^t_{n,i} =& \sum_{j=1}^{I_{n-1}} W^n_{i,j} y^t_{n-1,j} + b_i, \\ \label{eq:NeuronFNN}
y'^t_{n,i} =& f\left( x^t_i \right), \label{eq:NodeOutFNN}
\end{eqnarray}
\noindent and the structure of Eq. \ref{eq:NeuronFNN} connects neurons of layer $n$ exclusively to neurons in layer $n-1$.
 For generality, we focus on global coupling between each layer with equal weights.  
 The impact of inter-layer connectivity is discussed in Section \ref{sec:depth}. 
 The connections from layer $n-1$ to layer $n$ is determined by matrix $\mathbf{W}^n$ of size $I_{n-1}\times I_n$ and for global coupling $W^n_{i,j}=1/I_{n-1}$. 
 For sake of clarity we label $\mathbf{W}^2$ and $\mathbf{W}^N$ as input $\mathbf{W}^{\rm in}$ and output $\mathbf{W}^{\rm out}$ matrices, respectively.
 
For a FNN with such a topology, the shifting constant in the last layer is
\begin{equation}\label{eq:shift_const}
C_4=b+\alpha(b+\alpha(b+(0+b)))=b(1+2\alpha^2) + \alpha b,
\end{equation}
\noindent where $(0+b)$ is the signal coming from the first layer, and transformations $\alpha(b+\cdots)$ are made in each hidden layer.
 For the more general case of an unknown number of layers $N\ge 4$, the shifting constant is
\begin{equation}\label{eq:shift_const_general}
C_N=b(1+2\alpha^{N-2}) + b\sum\limits_{j=1}^{N-3}\alpha^j.
\end{equation}

\subsection{Correlated and uncorrelated noise}\label{sec:FNN_corr_uncorr_noise}

\begin{figure}[tbp]
\center{\includegraphics[width=1\linewidth]{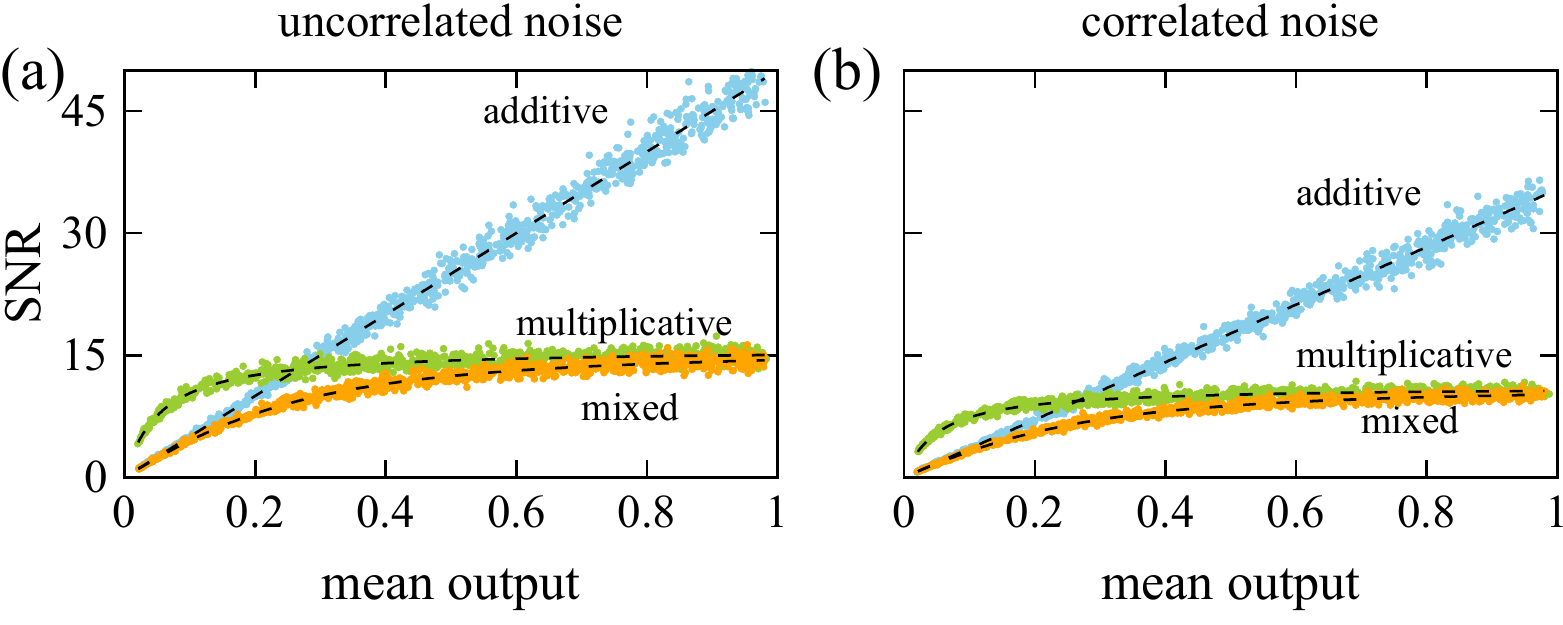}}
\caption{SNR dependences for uncorrelated noise (a) and correlated noise (b). Blue points correspond to only additive noise of intensity $D^U_A=D^C_A=10^{-4}$, green points indicate to only multiplicative noise with intensities $D^U_M=D^C_M=10^{-3}$, orange points match mixed noise with combination of both noise intensities. Parameters: $\alpha=1$, $b=0.02$.}\label{fig:FNN_corr_uncorr_noise}
\end{figure}

Within our framework, the state of each noisy neuron inside a hidden layer of a FNN exhibits two multiplicative and two additive noise sources, see Eq. (\ref{eq:one_neuron_noise}).
 Figure \ref{fig:FNN_corr_uncorr_noise} shows the numerically obtained SNR for exclusively uncorrelated noise in panel (a), and for exclusively correlated noise in panel (b).
 Both panels are identically scaled, and the detrimental impact of correlated noise compared to uncorrelated noise is clearly visible.
 However, corresponding dependencies do not differ qualitatively.
 Furthermore, there is little quailitative difference between a single node perturbed by noise, Fig. \ref{fig:SNR_one_node}, and a FNN's output, Fig. \ref{fig:FNN_corr_uncorr_noise}.
 From this comparison, we can conclude that additive, multiplicative and mixed noises have a very comparable effect on a FNN's and on a single noise neuron's output signal.
 The only notable difference is the contribution of multiplicative noise for small $E(y^t)$ values, which is due to shifting constant $C_4$.
  
In general, the overall SNR is reduced by the network.
 However, the accumulation of noise is very little: a single noisy neuron has less than twice the SNR compared to the collective of over 400 noisy neurons.
 In particular for the case of uncorrelated noise the reduction in SNR is moderate.

As the connection between the FNN layers is global, each neuron is driven by an averaged-state of the previous layer, a mean-field so to speak.
 This is where a network inherently aids keeping the influence of noise in check.
 Uncorrelated noise is different for each neuron of a preceding layer, and the influence of these different, uncorrelated noise-terms is efficiently averaged out by the network's connections.
 For correlated noise the situation is different.
 This term is identical for all neurons in preceding layers, averaging cannot curb the propagation of correlated noise. 
 This is an essential insight: in analogue neural networks an optimal network connectivity does not only address best the task it is proposed to solve, but also will depend on the network's very own noise properties.

Similar considerations also highlight the importance of neurons located in the first and last layers.
 The first layer spreads its noise across the entire system, and hence induces an additional form of correlated noise.
 The system's output, on the other hand, is the position where the computational results is presented to the outside world, and no averaging by network-connections is suppressing suppressing its uncorrelated noise.

\subsection{Signal to noise ratio}\label{sec:FNN_analytics}

The previous line of argument can be formulated analytically.
 Here we focus on the main results, the full derivation is given in Appendix~\ref{sec:Appendix_A}.

The mean value and variance of the the first layer are
\begin{eqnarray}
E(y^t_1)= E(y'^t_1)= u^t+b, \\\label{eq:FNN_mu_1}
\text{Var}(y^t_1)= \sigma^2_{add}+E^2(y^t_1)\cdot\sigma^2_{mult},\label{eq:FNN_var_1}
\end{eqnarray}
with $\sigma^2_{add}$ is the combined perturbation due to additive noise $\sigma^2_{add}=2D^U_A+2D^C_A$, and for multiplicative noise $\sigma^2_{mult}=2D^U_M+2D^C_M+4D^C_M D^U_M$.

Due to the here imposed symmetric topology, we find that 
\begin{equation}\label{eq:FNN_mu_n}
E(y^t_{n,i})=E(y'^t_{n,i})=E(y'^t_n)=\alpha(E(y'^t_{n-1})+b)
\end{equation}
This formula is valid for any hidden layer. 
 The variable $y'^t_n$ does not contain the noise of $n$th layer but has all the noise from previous layers. 
 The corresponding variance can be approximated according to
\begin{equation}\label{eq:FNN_var_n}
\begin{array}{c}
 \text{Var}(y'^t_n) \approx \alpha^2 \big[ 2D^C_A + 2D^C_M E^2(y'^t_{n-1}) +\\
(1+2D^C_M) \text{Var}(y'^t_{n-1}) \big] , \ \ 2<n<N.
\end{array}
\end{equation}
\noindent Equation (\ref{eq:FNN_var_n}) is used until $n=2$. 
 The variance for the 2nd layer is $ \text{Var}(y'^t_{2,n})=\alpha^2  \text{Var}(y^t_1)$. 
 Equations (\ref{eq:FNN_mu_n}) and (\ref{eq:FNN_var_n}) are recursively defined in function of the network's depth.
 From these relationships it becomes clear that the variance of hidden layers does not contain uncorrelated noise-contributions originating from the previous hidden layers.

Finally, the $N$th layer mean signal and variance, hence of the output signal are 
\begin{eqnarray}
E(y^t_N)= E(y^t_{N-1})+b \\\label{eq:FNN_mu_N}
\begin{array}{c}
\text{Var}(y^t_N) \approx \sigma^2_{add}+E^2(y^t_N)\cdot\sigma^2_{mult} + \\
(1+\sigma^2_{mult})\cdot \text{Var}(y'^t_N).\label{eq:FNN_var_N}
\end{array}
\end{eqnarray}
Equations (\ref{eq:FNN_var_1}) and (\ref{eq:FNN_var_N}) confirm that the first and last layer add both, correlated and uncorrelated noise to the output signal.

Based on these considerations, we arrive at the system's final SNR
\begin{equation}\label{eq:FNN_SNR_N}
\text{SNR}(y^t_N)=\frac{E(y^t_N)}{\big(\text{Var}(y^t_N)\big)^{1/2}},
\end{equation}
\noindent which only requires knowledge of individual component's noise properties.
 Figure \ref{fig:FNN_corr_uncorr_noise} shows analytically derived dependencies as dashed lines.
 The agreement between the numerical simulation and the analytical description is excellent.
 The noise of hardware neurons inside a network can be characterized in model systems, while the properties of input as well as the desired output signals are typically known for most tasks.
 The here derived analytical description is therefore of important predictive value for estimating hardware-linked performance limits in future analogue neural networks.


\section{Noise in recurrent networks\label{sec:RNN}}

\begin{figure}[tbp]
\center{\includegraphics[width=1\linewidth]{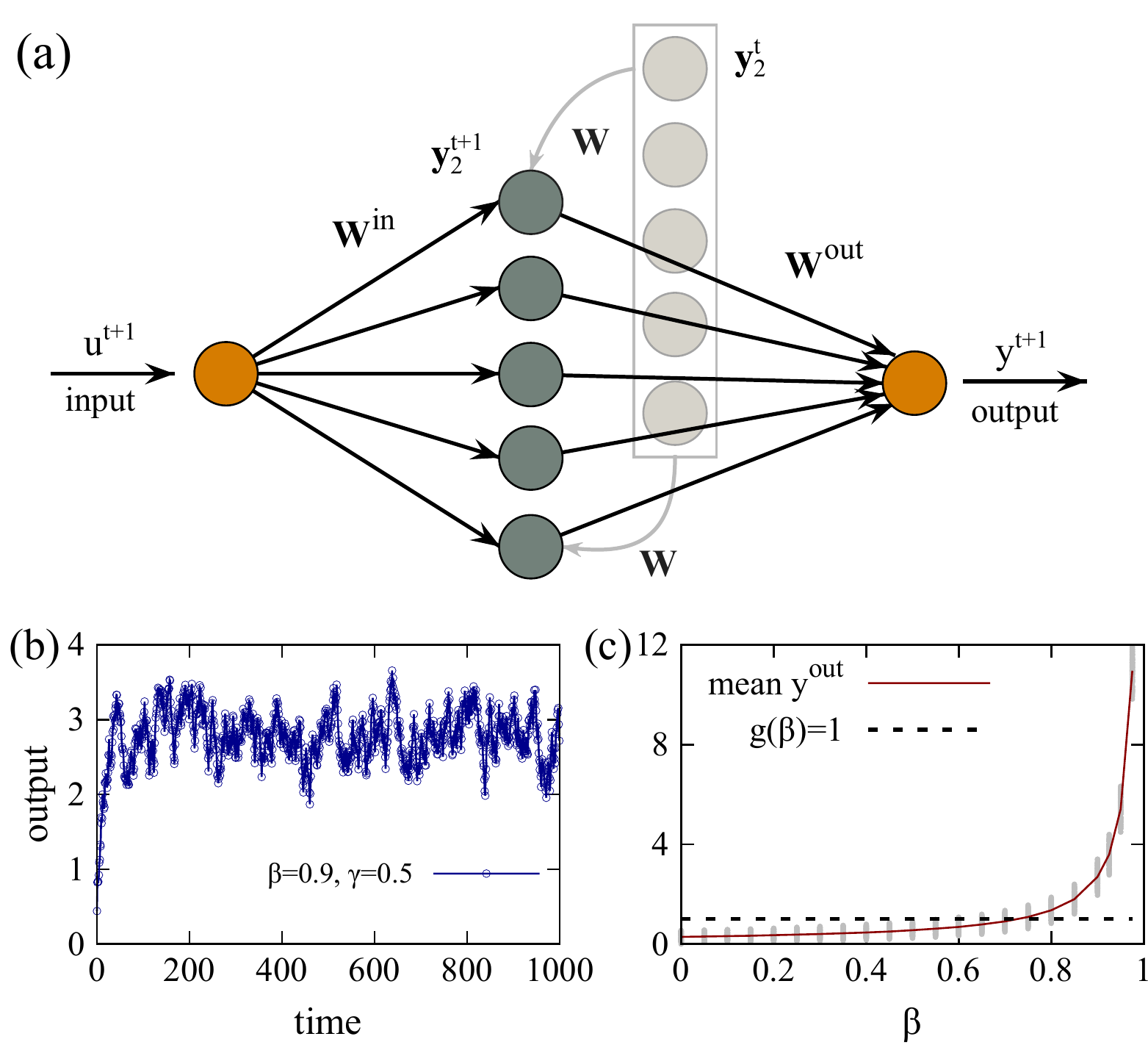}}
\caption[]{Schematic representation of simple recurrent neural network (1) consisting of input and output neurons (orange circles) and hidden (recurrent) layer (gray circles) of $I_2=200$ nodes. The panel (b) illustrates the output signal for $\gamma=0.5$, $\beta=0.9$. The panel (c) depicts shifts of output signal range (gray points) and focusing mean value (red line). The coupling is global, $\alpha=1$, $b=0.02$.}
\label{fig:RNN_scheme_realization}
\end{figure}

Following the previously developed techniques, we now turn to propagation of noise through RNNs.
 Figure \ref{fig:RNN_scheme_realization}(a) shows the general scheme of such a network.
 In our case, the network contains one input and one output neuron (orange circles), while the single hidden layer hosts $I_{2}=200$ neurons.  
 As for the case of the feed-forward network, the additional bias constant acts for all network nodes. 
 The evolution of neurons in the hidden-layer is given by
\begin{equation}\label{eq:RNN_network}
y'^{t+1}_{2,i}=f\big( \gamma\mathbf{W}^{\rm in} y^{t+1}_1  + \beta\mathbf{W} \mathbf{y}^t_2 + b \big),
\end{equation}
\noindent The output signal of our RNN is 
\begin{equation}\label{eq:RNN_yout}
y'^{t+1}_3=\mathbf{W}^{\rm out} \mathbf{y}^{t+1}_2 + b.
\end{equation}
\noindent Matrices $\mathbf{W}^{\rm in}$ and $\mathbf{W}^{\rm out}$ define  the connections between the hidden layer to the input and output neurons, respectively. 
 In contrast to FNNs, the system's state does not only depend on input signal $u^{t+1}$, but also on the hidden layer's state at previous times according to $\mathbf{W}$ in Eq. \ref{eq:RNN_network}.
 There is only one recurrent layer, so the index $n$ is not used for RNN.
 Parameters $\beta$ and $\gamma$ can be interpreted as balancing coefficient between the system's previous state $x^{t}$ or the current input signal $u^{t+1}$. 
 If $\beta=0$ and $\gamma=1$ the network has a FNN topology with a single hidden layer.
 For $\gamma=\beta=0.5$ the input signal and the recurrent signal have equal force. 

As in FNNs, it is important to consider the impact of constant signal accumulation in an RNN.
 However, the interplay between recurrence and input signal leads to a nonlinear modification of an RNN's output signal. 
 As the nodes are linear and we normalized all matrices to their largest eigenvalue, a FNN's output range depends only on the input signal's range. 
 In a RNN, this limit is fundamentally harder to obtain. 
 Figure \ref{fig:RNN_scheme_realization}(b) shows a RNN's output signal for $\alpha=1$, $\beta=0.9$ and $\gamma=0.5$, indicating the problem's temporal sensitivity via a short initial transient. 
 After this transient, the network's signal is dispersed around some mean value.
 However, due to the recurrence output range depends not only on the input signals amplitude range, but also on its temporal characteristic.
 As before, we drive the system with a random signal uniformly distributed within interval $u^{t}\in(0,\cdots,1)$. 
 Figure \ref{fig:RNN_scheme_realization}(c) shows the mean output signal as a red line for different strength of the network internal coupling $\beta$.

As expected, for $\beta$ approaching 1 the system destabilizes, and the offset quickly diverges towards infinity (with its pole at $\beta=1$).
 In a nonlinear system, this is the point were chaos due to sensitivity to initial conditions would start to arise.
 However, as we lack nonlinearity, resulting dynamics are not bound through nonlinear stretching and folding.
 However, we can approximate the amplitude-range of the output if the input signal is known. The mean value of the output signal $E(y^t_3)$ can be found for each time $t$. 
\begin{equation}\label{eq:RNN_mu_y}
\begin{array}{c}
E(y^t_3) = E(y'^t_2)+b, \ \ \ \text{where} \\ 
E(y'^t_2) = \alpha\Big( \gamma(u^t+b) + \beta E(y'^{t-1}_2) +b \Big).
\end{array}
\end{equation}
The full technique is given in Appendix B. Based on this technique, we obtain:
\begin{equation}
\begin{array}{c}\label{eq:RNN_range}
y_{min}=min\big( E(y^t_3) \big) \\
y_{max}=max\big( E(y^t_3) \big).
\end{array}
\end{equation}
They will be employed for normalizing output signal $y^t_3$ in order to keep it within the range $y^t_3\in(0,\cdots,1)$.

\subsection{Correlated and uncorrelated noise}

\begin{figure}[htbp]
\center{\includegraphics[width=1\linewidth]{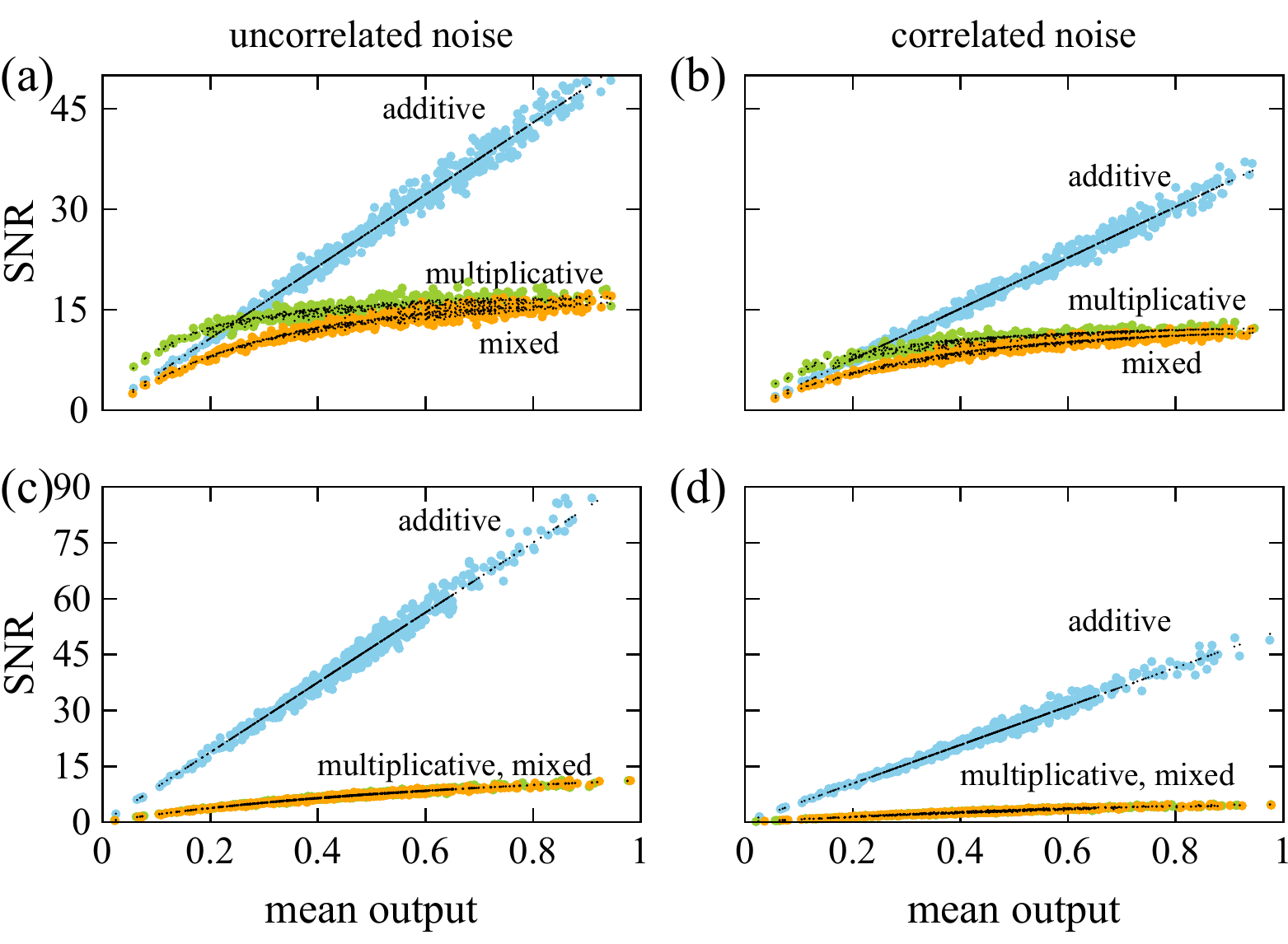}}
\caption[]{SNR dependences for uncorrelated noise (left panels) and correlated noise (right panels). Blue points correspond to only additive noise of intensity $D^U_A=D^C_A=10^{-4}$, green points indicate to only multiplicative noise with intensities $D^U_M=D^C_M=10^{-3}$, orange points match mixed noise with combination of both noise intensities. Top panels were prepared for $\beta=0.5$, bottom panels correspond to $\beta=0.9$. Other parameters are: $\gamma=0.5$, $I_2=200$, $\alpha=1$.}
\label{fig:RNN_SNR}
\end{figure}

Figure \ref{fig:RNN_SNR}(a) and (b) show the SNR for $\beta=\gamma=0.5$ for exclusively uncorrelated and correlated noise, respectively.
 Noise intensities are identical to the once of the FNN-section, and the resulting SNR has overall similar properties and scale as the one found for the 4-layer FNN.
 Again, correlated noise significantly reduces the system's performance when compared to uncorrelated noise, which indicates that the temporal averaging in an RNN suppresses uncorrelated noise in a similar fashion as in deep FNNs.

Increasing $\beta$ confirms this general observation, see Fig. \ref{fig:RNN_SNR}(a) and (d) for exclusively uncorrelated and correlated noise at for $\beta=0.9$ and $\gamma=0.5$, respectively.
 The increased contribution of the recurrent signal has multiple consequences. 
 On one hand $\beta$ improves the additive SNR, but then on the other hand significantly reduces the multiplicative SNR.
 Ultimately, the output's SNR is dominated by the latter contribution, and hence increasing $\beta$ increases the RNN's susceptibility to noisy neurons.

\subsection{\label{sec:RNN_analytics}Noise in recurrent networks}

Here, we face the same challenge as when deriving the RNN's output amplitude range: the analytical SNR description requires some knowledge of the input signal's temporal nature.
 As before we will focus on the final form of the equations; the complete derivation can be found in Appendix \ref{sec:Appendix_B}. 

For a noisy RNN, the mean value is (\ref{eq:RNN_mu_y}). The noise-induced variance of its output signal is given by
\begin{equation}\label{eq:RNN_var_y}
\begin{array}{l}
 \text{Var}(y^t_3) = \sigma^2_{add} + \sigma^2_{mult} E^2(y^t_3) + (1+\sigma^2_{mult})\times \\
 \big[ 2D^C_A + 2D^C_M E^2(y'^t_2) + (1+2D^C_M)  \text{Var}(y'^t_2)  \big].
\end{array}
\end{equation}
\noindent Variables $\sigma^2_{add}$ and $\sigma^2_{mult}$ are identical to the once previously introduced in Sec.~\ref{sec:FNN_analytics}.
 Furthermore, $y'^t_2=y'^t_{2,i}$ contains the averaged signal passed on from the recurrent layer of previous time ($t-1$) to the output neuron for the case of global coupling.
 Its variance is
 \begin{equation}\label{eq:RNN_y2_var}
\begin{array}{c}
 \text{Var}(y'^t_2)\approx \alpha^2\gamma^2  \text{Var}(y^t_1) + \alpha^2\beta^2 \big[ 2D^C_A + \\
2D^C_M E^2(y'^{t-1}_2) + (1+2D^C_M)  \text{Var}(y'^{t-1}_2)  \big].
\end{array}
\end{equation}
with mean value
\begin{equation}\label{eq:RNN_y2_mean}
E(y'^t_2)=E(y^t_{2,i}) = \alpha \big( \gamma E(y^t_1) + \beta E(y'^{t-1}_2) +b  \big) \\
\end{equation}
Again the uncorrelated noise comes only from the input and output neurons.
 The nodes from the recurrent layer transmit only their correlated noise. It means that, as with FNN, the main noise effect comes from the input and output neuron.
 Equations (\ref{eq:RNN_mu_y})--(\ref{eq:RNN_y2_mean}) demonstrate good correspondence with the numerical simulation (Fig.~\ref{fig:RNN_SNR} black points).



\section{\label{sec:coupling}Noise propagation versus connectivity}

Previously we considered networks were connectivity in all matrices was 100$~\%$ and uniform.
This enabled a clean derivation of analytical models, however, it does not correspond to typical topology-statistics of neural networks.
In the following sections we will replace the fully connected matrices by matrices consisting of random entries and with a certain fraction of non-zero connections, i.e. connectivity.

\subsection{Connectivity in FNNs}\label{sec:FNN_coupling}

We will again focus on the case of a deep FNN with a single input and output neuron, $I_1=I_4=1$, and hosting with $I_2=I_3=200$ neurons in the hidden layers.
Matrices $\mathbf{W}^{in}\in\mathbb{R}^{1\times I_2}$ and $\mathbf{W}^{out}\in\mathbb{R}^{I_3\times 1}$ determine the system's connection to the in and output, respectively.
Coupling between hidden layers is given by $\mathbf{W}^{\rm 3}\in\mathbb{R}^{I_2\times I_3}$.
All matrices have a percentage of $\rho$ non-zero entries, which are drawn from a random distribution.
Finally, matrices are normalized to their largest eigenvalue.
For $\rho=100~\%$ we would again obtain global coupling, however now according to random weight distributions.

\begin{figure}[h] 
\center{\includegraphics[width=1\linewidth]{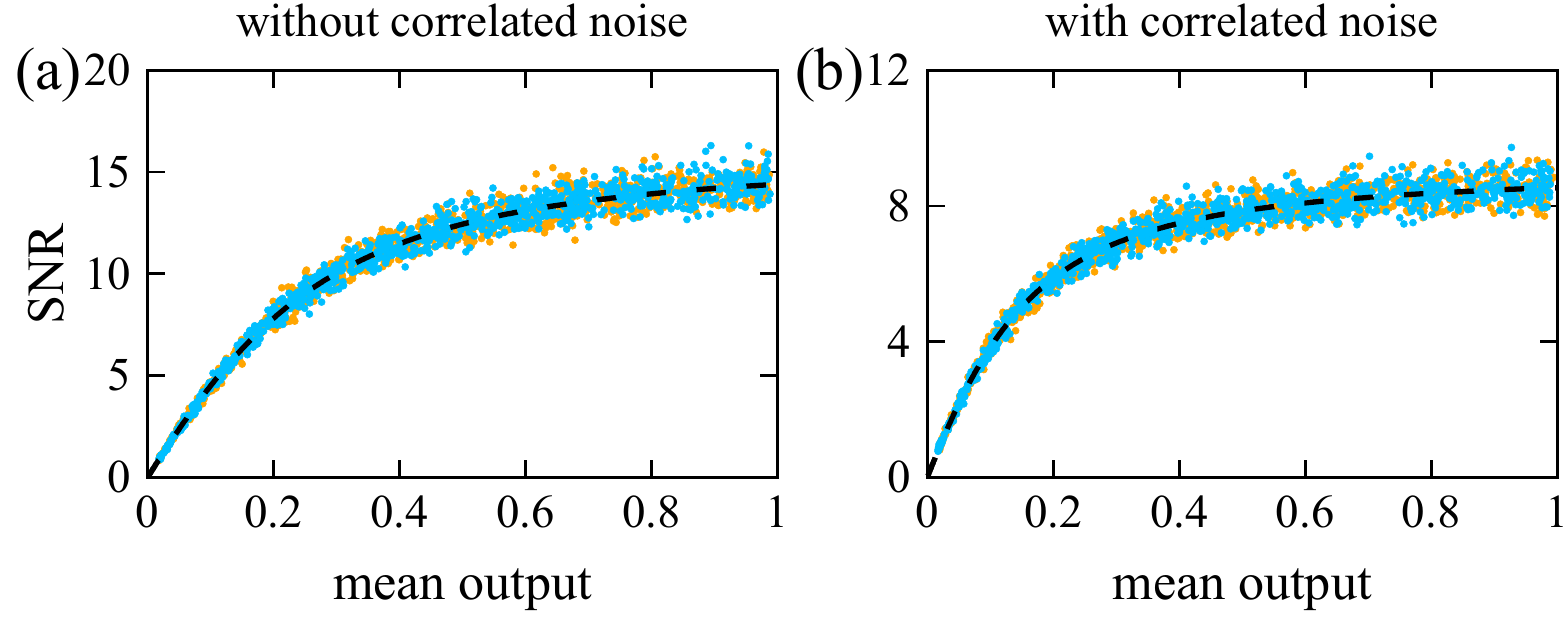}}
\caption{Signal-to-noise ratio in the last layer of FNN for the case with multiplicative correlated noise of intensity $D^C_M=10^{-3}$ (b) and without it (a). Other parameters are $D^U_A=10^{-4}$, $D^U_M=10^{-3}$. Blue points correspond 100\% connection and orange ones to 1\% connection between hidden layers. Both dependences are almost the same.}\label{fig:FNN_coupling_hidden}
\end{figure}

Figure \ref{fig:FNN_coupling_hidden} shows the SNR for uncorrelated and correlated noise in panels (a) and (b).
The orange (blue) data have been obtained using $\rho=1~\%$ ($\rho=100~\%$) connectivity of $\mathbf{W}^{\rm 3}$, i.e. between the two hidden layers, while $\mathbf{W}^{\rm 2}$ and $\mathbf{W}^{\rm 4}$ were fully connected.
The dashed line was obtained based on the analytical SNR description based on Eqs. (\ref{eq:FNN_mu_N})-(\ref{eq:FNN_SNR_N}), crucially dervide under the assumption of symmetric and full connectivity.
From the data it is clear that the SNR's dependency on the connectivity between both hidden layers is very weak.
Furthermore, for correlated as well as uncorrelated noise the analytical SNR description perfectly agrees with the numerically obtained data.
This leads to an interesting conclusion: attenuation of uncorrelated noise is already established by averaging tue to input and readout matrices $\mathbf{W}^{\rm 2}$ and $\mathbf{W}^{\rm 4}$.

\begin{figure}[h] 
\center{\includegraphics[width=1\linewidth]{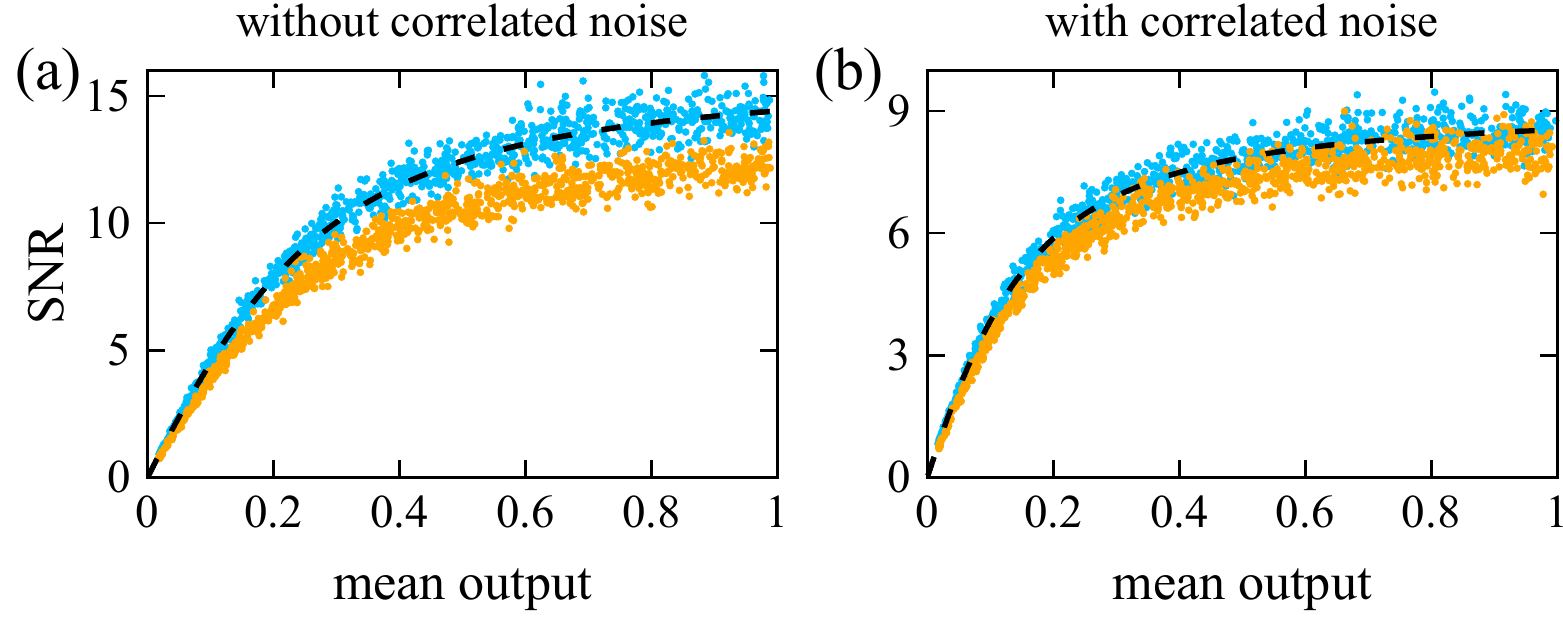}}
\caption{Signal-to-noise ratio in the last layer of FNN for the case with multiplicative correlated noise with intensity $D^C_M=10^{-3}$ (b) and without it (a). Other parameters are $D^U_A=10^{-4}$, $D^U_M=10^{-3}$. Blue points correspond 100\% connection and orange ones to 1\% connection in the output matrix $\mathbf{W}^{\rm out}$. The matrix between hidden layers has 1\% connection.}\label{fig:FNN_coupling_out}
\end{figure}

For this reason, we now turn our attention to the SNR's dependency upon the readout connectivity.
The corresponding SNR, see Fig.~\ref{fig:FNN_coupling_out} was obtained for $\rho=1~\%$ of $\mathbf{W}^{\rm 3}$ and $\rho=100~\%$ of $\mathbf{W}^{in}$, while orange (blue) data corresponds to $\rho=1~\%$ ($\rho=100~\%$) connectivity for $\mathbf{W}^{out}$, respectively.
Only in the case of a final layer connectivity of $\rho=1~\%$ we obtain a reduction of the SNR-function for uncorrelated noise, hence a reduced suppression of such.
Such a low final layer connectivity is unlikely for most applications. 

\subsection{Coupling in RNN}\label{sec:RNN_coupling}

Guided by the previous FNN-results we can immediately focus on the connectivity of the RNN's readout layer.
Crucially, we have confirmed that, as for the FNN, $\rho$ of the RNN's hidden layer $\mathbf{W}^{out}$ has negligible impact upon the system's SNR.
However, there is a fundamental difference when discussing the RNN's connectivity.
Besides the percentage of non-zero connections $\rho$, parameter $\beta$ greatly influences the system's sensitivity as well as its connection topology, and is therefore included in our considerations.

\begin{figure}[h] 
\center{\includegraphics[width=1\linewidth]{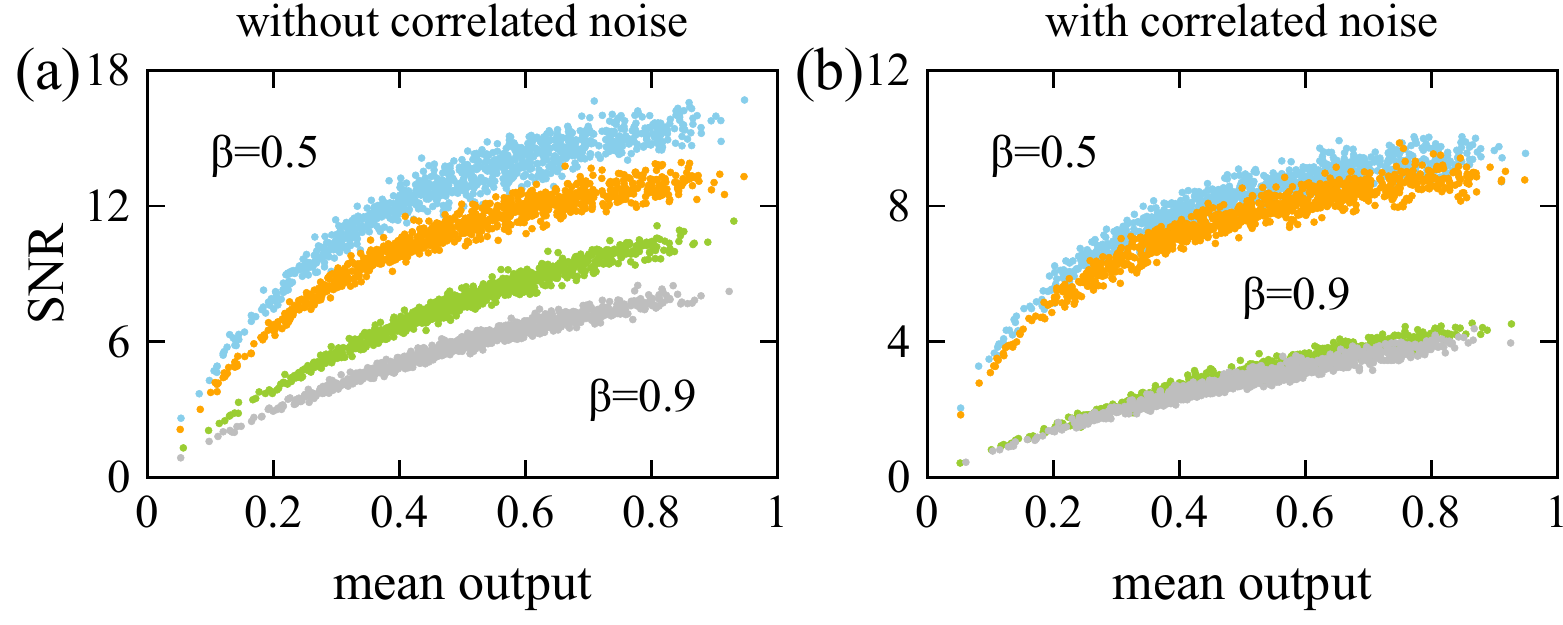}}
\caption{Signal-to-noise ratio in RNN for the case with multiplicative correlated noise with intensity $D^C_M=10^{-3}$ (b) and without it (a). Other parameters are $D^U_A=10^{-4}$, $D^U_M=10^{-3}$. Blue points ($\beta=0.5$) and green points ($\beta=0.9$) correspond 100\% connection and orange ($\beta=0.5$) and gray ($\beta=0.9$) ones are for 1\% connection in the output matrix $\mathbf{W}^{\rm out}$. The connectivity of the hidden (input) layer is 1~$\%$ (100~$\%$).}\label{fig:RNN_coupling_out}
\end{figure}

For each noise-type and final layer connectivity we investigate two $\beta$-values.
In Fig.~\ref{fig:RNN_coupling_out} we show the SNR for $\beta=0.5$ and final layer connectivity $\rho=1~\%$ ($\rho=100~\%$) as orange (blue) data.
Data for $\beta=0.9$ and final layer connectivity $\rho=1~\%$ ($\rho=100~\%$) is shown in gray (green).
For uncorrelated noise, panel (a), we find that for both, connectivity as well as $\beta$ have a strong impact.
Again, higher connectivity suppresses propagation of uncorrelated noise, while larger $\beta$'s increase the system's sensitivity.
As expected, we find very little impact of the final layer's connectivity upon the suppression of correlated noise for $\beta=0.5$.
For $\beta=0.9$ we find that connectivity has no positive effect upon correlated noise suppression; a $\rho$ of the final layer beyond $\sim10~\%$ is sufficient for that purpose.

\begin{figure}[h] 
\center{\includegraphics[width=1\linewidth]{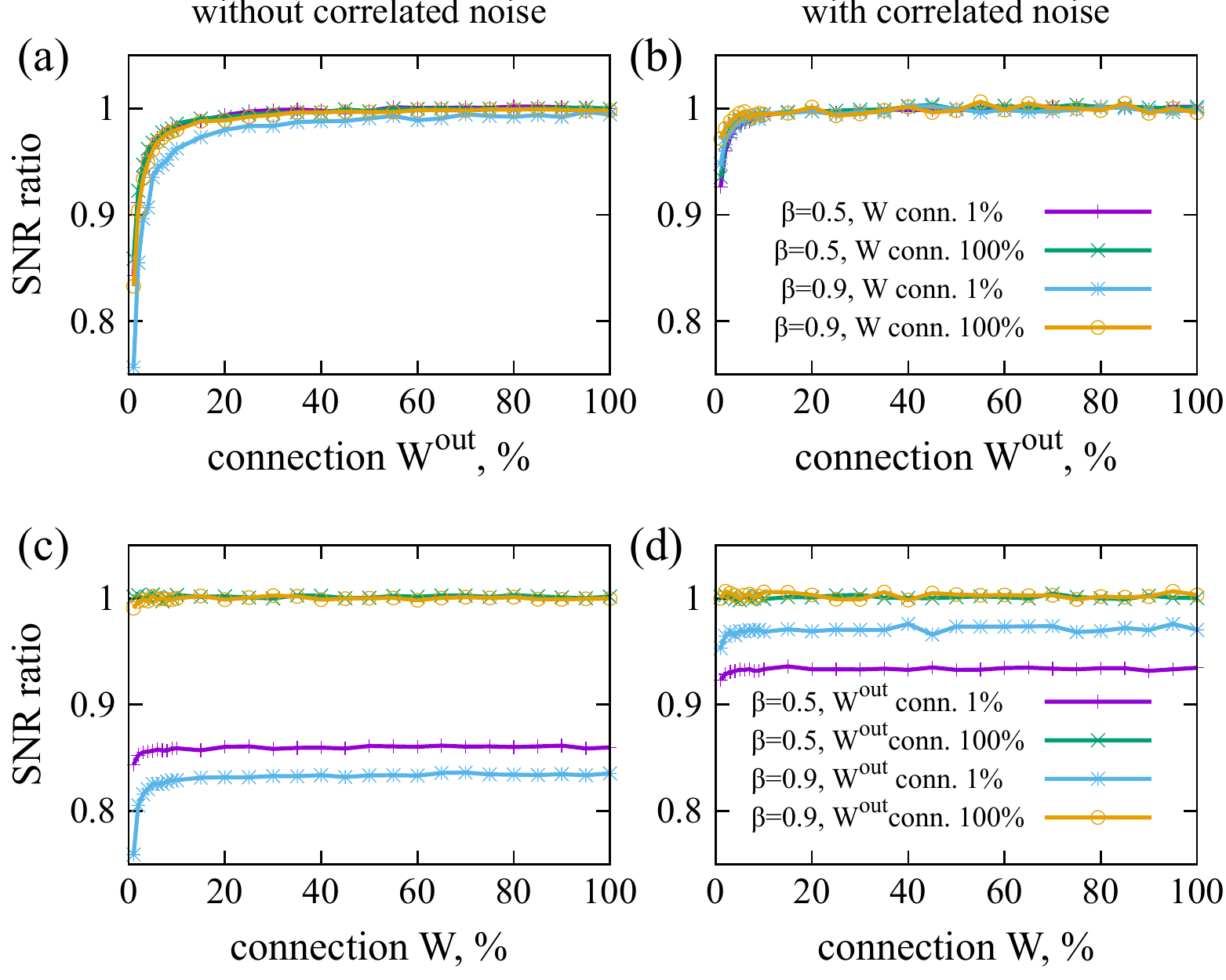}}
\caption{SNR ratio for RNN. SNR dependences are normalized to SNR for corresponding $\beta$ and nose intensities but for global coupling in all the matrices.}\label{fig:RNN_SNR_ratio_conn}
\end{figure}


\section{\label{sec:depth} Impact of depth}

\begin{figure}[tbp] 
\center{\includegraphics[width=1\linewidth]{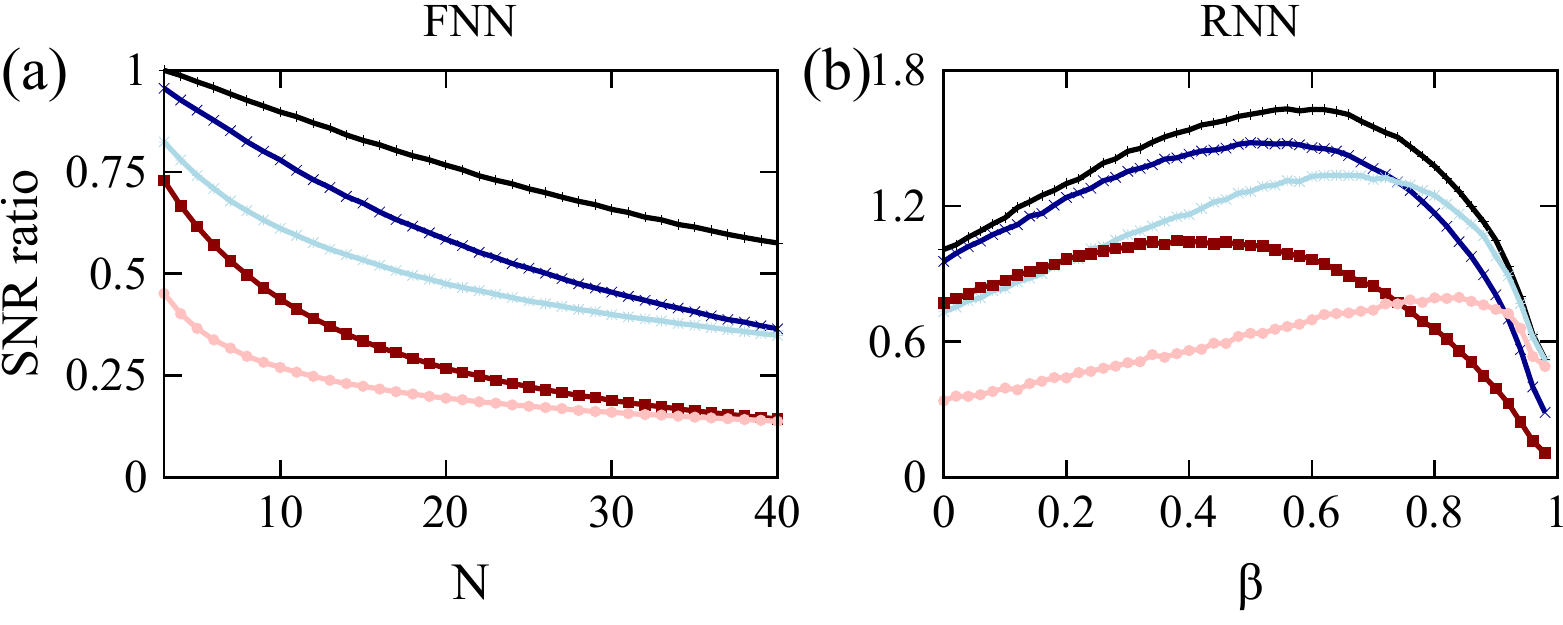}}
	\caption{SNR ratio for different number of layers in FNN (a) and different recurrency impact $\beta$ in RNN (b). The color scheme is the same for both panels. Black lines correspond to the case without correlated noise. Other lines correspond to different correlated noise intensities: dark blue -- $D^C_M=10^{-4}$, dark red -- $D^C_M=10^{-3}$, light blue -- $D^C_A=10^{-4}$, light red -- $D^C_A=10^{-3}$. Fixed parameters are $D^U_A=10^{-4}$, $D^C_M=10^{-4}$, $D^U_M=10^{-3}$, $\alpha=1$, $b=0.02$. }\label{fig:depth}
\end{figure}

The final topological consideration in our work will be the impact of a FNN's and RNN's depth upon the SNR.
 The feed-forward network in the simplest case has only one hidden layer ($N=3$), and its SNR for exclusively uncorrelated noise ($D^U_M=10^{-3}$, $D^U_A=10^{-4}$) will serve as the reference for deeper FNNs.
 Figure~\ref{fig:depth}(a) illustrates the SNR's dependency upon the number of layers for five cases: without correlated noise (black data), with additive correlated noise (light red ($D^C_A=10^{-4}$) and light blue ($D^C_A=10^{-3}$) data) and with multiplicative correlated noise (dark red ($D^C_M=10^{-4}$) data and dark blue ($D^C_M=10^{-3}$) data).
 In general, the FNN is more resilient against multiplicative noise, and while the FNN's depth certainly has an impact, it requires approximately 20 layers for the SNR to be reduced to half of the single hidden layer system.

Figure~\ref{fig:depth}(b) illustrates the RNN's dependency upon its internal coupling strength $\beta$ for the same five configurations of noise.
 The SNR reference was obtained for the case without correlated noise and $\beta=0$, black data.
 As mentioned before, the internal coupling strength can be linked to the depth of a corresponding mapping of the RNN upon an FNN via the unfolding in time technique.
 For $\beta=1$ the depth of such a system becomes infinity, and it is hence not surprising that for this case the SNR drops to zero for all noise-configurations.
 Here we would like to point out a difference we expect for the case of nonlinear RNNs.
 For $\beta>1$ the nonlinearity results in characteristic dynamical regimes, and for the case of periodic orbits we expect that the system will still preserve a large degree of its noise-resilience \cite{Marquez2018}.
 Apart from the SNR when approaching the critical value of $\beta=1$ we find that the system has an ideal operational point for maximum noise mitigation.
 The particular value of the optimal $\beta$ depends on the type and amplitude of noise.


\section{\label{sec:Startegies}Strategies for noise mitigation}

In the previous sections we have continuously encountered two essential aspects of a analogue neural networks sensitivity to noise.
 The first is that uncorrelated noise can efficiently be suppressed through averaging by the numerous network's connections.
 And second is that the global SNR dependencies of FNN's and RNN's share strong similarities.
 We will now use these features for devising first strategies for boosting neural network noise mitigation.

\begin{figure}[tbp] 
\center{\includegraphics[width=1\linewidth]{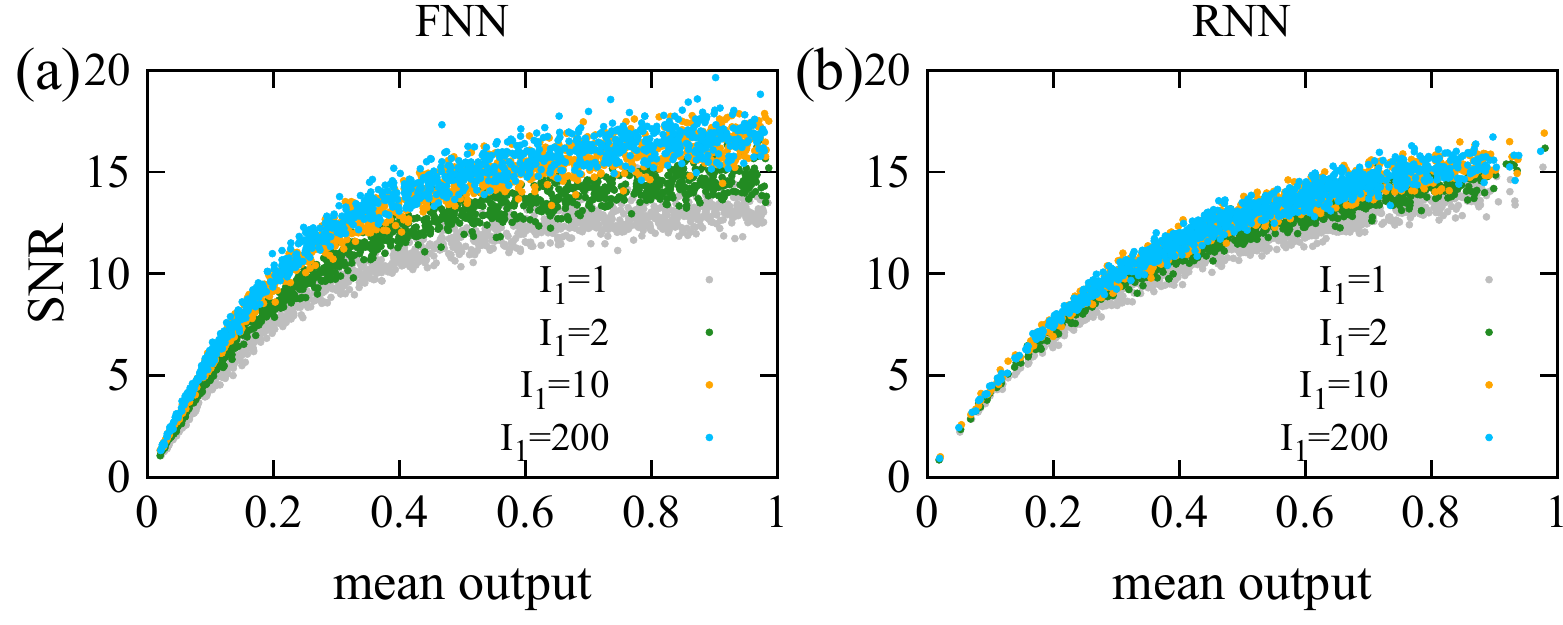}}
\caption{Signal-to-noise ratio of the output signal in FNN (a) and RNN (b) for the case different numbers of nodes in the input layer and noiseless neurons in the final readout layer. Parameters are $D^U_A=10^{-4}$, $D^C_M=10^{-4}$, $D^U_M=10^{-3}$, $\alpha=1$, $b=0.02$, $\beta=0.5$. }\label{fig:I1_with_noise4}
\end{figure}

The first argument in the previous paragraph leads the way to an intuitive strategy.
 The fact that the first layer neuron acts as drive for all other neurons morphs this particular element into the ultimate source of globally correlated noise.
 We therefore attempt to de-correlate this impact and increase the number of input neurons where each of the input neurons receives the same input signal.
 Figure \ref{fig:I1_with_noise4} illustrates the system's SNR for identical noise conditions for different $1 \leq I_1 \leq 200$. 
 Panel (a) shows the impact of this strategy upon a FNN, panel (b) for a RNN ($\gamma=\beta=0.5$). 
 For the FNN, the impact of increasing $I_1$ is significant and improve the SNR by $\sim30~\%$ for $I_1 \gtrsim 10$.
 For the RNN we do not find significant impact.
 
\begin{figure}[tbp] 
\center{\includegraphics[width=1\linewidth]{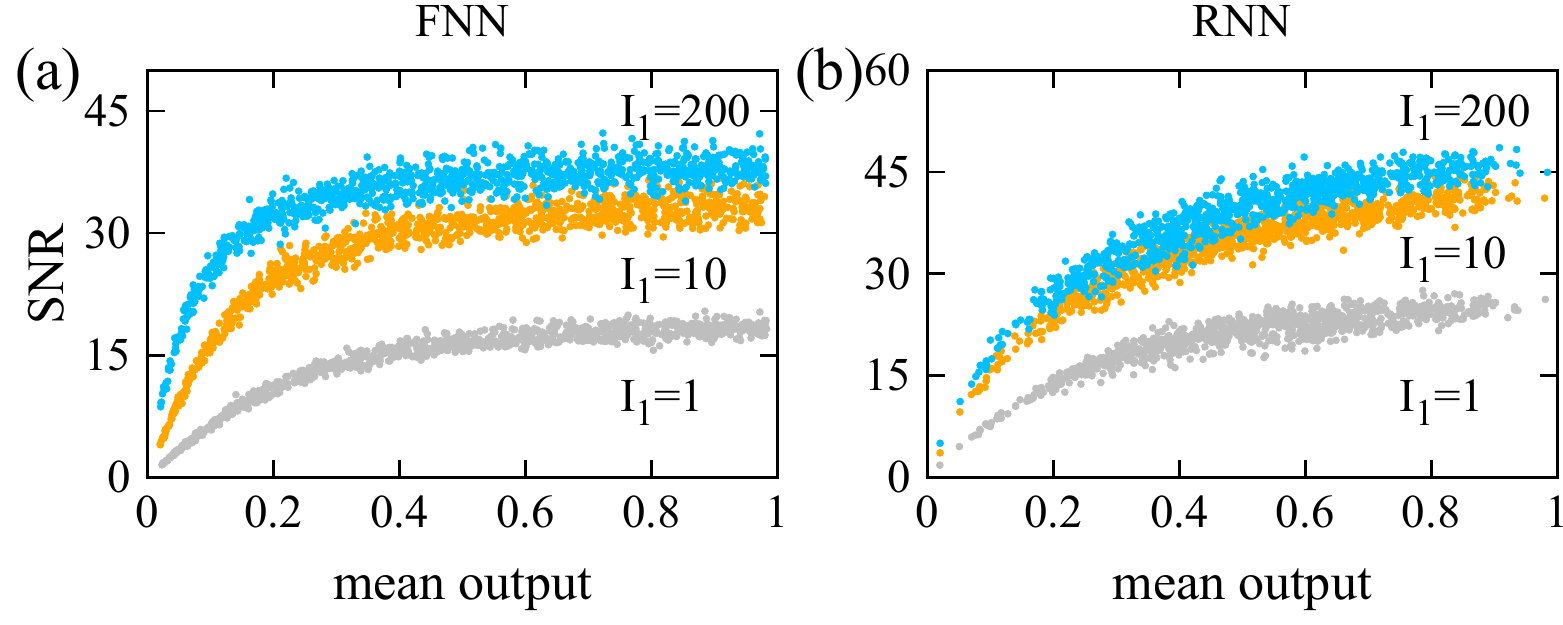}}
\caption{Signal-to-noise ratio of the output signal in FNN (a) and RNN (b) for the case different numbers of nodes in the input layer and turned off noise in the output layer. Parameters are $D^U_A=10^{-4}$, $D^C_M=10^{-4}$, $D^U_M=10^{-3}$, $\alpha=1$, $b=0.02$, $\beta=0.5$. }\label{fig:I1_no_noise4}
\end{figure}

The relevance of the last layer is due to the fact that fundamentally there its no further averaging taking place after.
 The most immediate conclusion is therefore to place substantial effort on low-noise last layer neurons in the fabrication and design of an analogue neural networks.
 In the case of the systems as in \cite{Bueno2018}, this corresponds to the selection of a low-noise detector in the output.
 Figure \ref{fig:I1_no_noise4} shows the great effectiveness of this strategy.
 In both cases, for the FNN, panel (a), and RNN, panel (b) ($\gamma=\beta=0.5$), the benefit upon the system's SNR is significant when combined with multiplexing of the first layer input neurons. 
 Based on noise-less readout neurons and duplicating the input-neuron to around 10 copies approximately doubles the system's SNR.
 Most importantly, we have also analysed the situation for only uncorrelated noise.
 There, the SNR can be increased close to 10 times following this strategy.

In some types of networks, it is impossible to alter the neuron type. Then the idea of turned off noise can be transformed to the next way. Previously the feed-forward network with four layers of nodes $I_1$, $I_2$, $I_3$, $I_4$ with $I_4=1$ has been considered. Now we remove the last layer and read out all the signal from each node of the 3rd layer. Then we average this set of signals over the number of nodes in the third layer $I_3$ and get almost the same effect of noise reduction. The same can be made for RNN.

\section{\label{sec:Conclusion}Conclusion}

To conclude, we have analysed the interplay between noise and network connections in great detail.
 We have considered a large variety of networks topologies and parameters are have analysed their output susceptibility to noisy neurons.
 We find that the in general the networks are quite resilient towards such noise units.

The principle results of our work are confirmed by analytical descriptions.
 They identify the fundamental laws with govern the propagation of noise across future hardware implementations.
 Based on the gain insight identify noise which is correlated across populations of neurons together with the noise present in input and output neurons as the main nuisance in such systems.
 Our work identifies the fundamental importance of stability in global parameters, for example in the stability of a analogue neural network's power supply.
 Individual neurons, on the other hand, can be of lesser quality as their local noise will largely be averaged out by the network itself.
 
We directly leverage this insight and propose noise mitigation strategies which have a great effect.
 Under good conditions this approach can results into an SNR improvement by up to one order of magnitude.

Our work focused on networks of linear components.
 By that it highlighted the fundamental interaction between noisy neurons and network connections.
 However, we presented a large fraction of our results as a SNR dependency against the average output amplitude.
 This allows for a first, heuristic mapping of our findings onto nonliear system's for which the nonlinearity is known.
 An important task is now to include such nonlinear noisy neurons and to understand their role inside the noisy orchestra in detail.

\begin{acknowledgments}

The authors acknowledge the support of the Region Bourgogne Franche-Comt\'{e}.
This work was supported by the EUR EIPHI program (Contract No. ANR-17-EURE- 0002), the BiPhoProc project (Contract No. ANR-14-OHRI- 0002-02), by the Volkwagen Foundation (NeuroQNet).
N.S. is supported by Vernadski scholarship of the French Government. 
X.P. has received funding from the European Union’s Horizon 2020 research and innovation programme under the Marie Sklodowska-Curie grant agreement No. 713694 (MULTIPLY).

\end{acknowledgments}

\appendix

\section{Analytical prediction of SNR in FNN}\label{sec:Appendix_A}

In this section the analytical prediction of SNR in FNN is obtained. 
 The next formulas are valid only for three main conditions: (i) the coupling between layers is global; (ii) the number of nodes in hidden layer is large $I_n\gg 1$; (iii) all noise sources $\xi$ have zero mean value and $1$ variance. 
 The statistical characteristics of noise are controlled by corresponding noise intensities.

In the present manuscript we consider the case when the function $f(x)=\alpha x$ affects only in hidden layers. 
 Following the nomenclature of Sec. III the evolution equation for the first layer $y'^t_1$ and the value after noise influence $y^t_1$ are described by
\begin{equation}\label{eq:AppA_y1}
\begin{array}{l}
y'^t_1=u^t+b \\
y^t_1=y'^t_1\cdot(1+\sqrt{2D^C_M}\xi^C_{M1})(1+\sqrt{2D^U_M}\xi^U_{M1}) + \\
 \sqrt{2D^C_A} \xi^C_{A1} + \sqrt{2D^U_A} \xi^U_{A1}.
\end{array}
\end{equation}
The expected value for the first layer is $E(y^t_1)=u^t+b$. 
 The corresponding variance can be found using three main rules of random variables algebra: 
 (i) the variance of sum is $Var(\xi+\eta)=Var(\xi)+Var(\eta)$, where $\xi$ and $\eta$ are random variables which are not correlated to each other. 
 (ii) the variance of their multiplication is $Var(\xi\cdot\eta) = (E^2(\eta)+Var(\eta))\cdot Var(\xi) + E^2(\xi)Var(\eta)$. 
 (iii) the multiplication by some constant has the variance $Var(c\xi)=c^2 Var(\xi)$. 
 Then the variance in the first layer is:
\begin{equation}\label{eq:AppA_y1_var}
Var(y^t_1) = \sigma^2_{add} + \sigma^2_{mult} (u^t+b)^2,
\end{equation}
where  $\sigma^2_{add}=2D^C_A+2D^U_A$ shows the variance coming from only additive noises and $\sigma^2_{mult}=2D^C_M+2D^U_M+4D^C_M D^U_M$ is the variance of the multiplier with multiplicative noises after $y'^t_1$.

The general noisy equation for each layer $n$ has a common form
\begin{equation}\label{eq:AppA_yn_noise}
\begin{array}{r}
y^t_{n,i} = y'^t_{n,i}\cdot(1+\sqrt{2D^C_M}\xi^C_{M1})(1+\sqrt{2D^U_M}\xi^U_{M1}) + \\
 \sqrt{2D^C_A} \xi^C_{A1} + \sqrt{2D^U_A} \xi^U_{A1}, \ \ \ 1\le n\le N
\end{array}
\end{equation}
and depends on the state without noise $y'^t_{n,i}$. 
 The coupling is global between hidden layers. 
 Then each $i$th neuron of  the $n$th layer receives the same signal from the previous one. 
 Therefore the index $i$ can be neglected in the equation for $y'^t_{n,i}$:
\begin{equation}
\begin{array}{r}
y'^t_n = y'^t_{n,i} = \alpha \big[ \frac{1}{I_{n-1}}\cdot \sum^{I_{n-1}}_{j=1} y^t_{n-1,j} +b \big] = \\
\alpha\big[  b + \sqrt{2D^C_A}\xi^C_{An-1} + y'^t_{n-1} (1+\sqrt{2D^C_M}\xi^C_{Mn-1})\cdot \\
\big(1+ \frac{1}{I_{n-1}}\sum^{I_{n-1}}_{j=1} \sqrt{2D^U_M}\xi^U_{Mn-1,j}  \big)    + \\
\frac{1}{I_{n-1}}\sum^{I_{n-1}}_{j=1} \sqrt{2D^U_A}\xi^U_{An-1,j}    \big] .
\end{array}
\end{equation}
Roughly speaking, the variance of the sum divided by $I_n$ equals to the sum of variances divided by $I^2_n$. 
 At $I_n\gg 1$ these components are much smaller than the rest, so they can be ignored. 
 They have no impact on expected values and variances.
\begin{equation}\label{eq:AppA_yn}
y'^t_n \approx \alpha \big[  b + y'^t_{n-1} \big(1+\sqrt{2D^C_M}\xi^C_{Mn-1}\big) + \sqrt{2D^C_A}\xi^C_{An-1}   \big] .
\end{equation}
It is clearly seen that the equation (\ref{eq:AppA_yn}) contains only correlated noise. 
 Only this type of noise goes from the hidden layers to the output signal. 
 The expected value for hidden layer is
\begin{equation}\label{eq:AppA_yn_mean}
E(y'^t_n) = E(y^t_{n,i}) = \alpha (b+E(y^t_{n-1})), \ \ 1<n<N
\end{equation}
The corresponding variance is
\begin{equation}\label{eq:AppA_yn_var}
Var(y'^t_n) \approx \alpha^2 \big[ 2D^C_A + 2D^C_M E^2(y'^t_{n-1}) + (1+2D^C_M)Var(y'^t_{n-1})  \big] .
\end{equation}
The equations (\ref{eq:AppA_yn}), (\ref{eq:AppA_yn_var}) are valid only for the hidden layers $2<n<N$. 
 The second layer has a small difference because the previous $1$st layer has only one node and there is no average over $I_1$. 
 Both correlated and uncorrelared noise from the first layer propagates through the network.

\begin{equation}\label{eq:AppA_y2}
y'^t_2 = y'^t_{2,i} = \alpha(y^t_1+b)
\end{equation}
\begin{equation}\label{eq:AppA_y2_var}
Var(y'^t_2) = \alpha^2 \cdot Var(y^t_{1}).
\end{equation}
The expected value in the 2nd layer is described by (\ref{eq:AppA_yn_mean}). The state equation of the last $N$th layer is
\begin{equation}\label{eq:AppA_yN}
y'^t_N = \frac{1}{I_{N-1}}\cdot \sum^{I_{N-1}}_{j=1} y^t_{N-1,j} +b .
\end{equation}
And the corresponding variance can be calculated using (\ref{eq:AppA_yn}) but without $\alpha$. The noisy equation (\ref{eq:AppA_yn_noise}) occurs also for $n=N$. Then the variance for the output signal is
\begin{equation}\label{eq:AppA_yN_var}
\begin{array}{r}
Var(y^t_N)\approx  \sigma^2_{add} + E^2(y^t_N)\sigma^2_{mult} + (1+\sigma^2_{mult})\cdot Var(y'^t_N) = \\
\sigma^2_{add} + E^2(y^t_N)\sigma^2_{mult} + (1+\sigma^2_{mult})\big[ 2D^C_A + \\
2D^C_M E^2(y'^t_{N-1}) + (1+2D^C_M)Var(y'^t_{N-1}) \big].
\end{array}
\end{equation}
The last term with $[\cdots]$ contains the noisy impact from the previous (N-1)th layer. 
 The variable $Var(y'^t_{N-1})$ can be replaced by the same term but with noise from the (N-2)th layer and so on until $Var(y'^t_2)$. 
 The variance for the second layer is calculated using (\ref{eq:AppA_y2_var}). 
 In the end the final variance of the output signal contains correlated and uncorrelated noise from the first and the last layer and only correlated noise from hidden layers.

The expected value of the output signal is
\begin{equation}\label{eq:AppA_yN_mean}
E(y^t_N) = E(y'^t_N)= E(y'^t_{N-1})+b.
\end{equation}
The signal-to-noise ratio can be found as follows:
\begin{equation}\label{eq:AppA_SNR}
SNR(y^t_N) = \frac{E(y^t_N)}{Var\big(y^t_{N-1}\big)^{1/2}}.
\end{equation}

\section{Analytical prediction of SNR in RNN}\label{sec:Appendix_B}

This section is devoted to the prediction of SNR dependence in the recurrent network. Three conditions are used here as for FNN: (i) global coupling between layers; (ii) the number of nodes in the recurrent layer is large $I_2\gg 1$; (iii) the input and output layer has only one node.

The structure of RNN is very similar to FNN and the noise has the same affect. So some of the equations will be the same as in Appendix \ref{sec:Appendix_A}.

The node in the first layer before the noise influence is described by the equation for $y'^t_1$ and by $y^t_1$ after it.
\begin{equation}\label{eq:AppB_y1}
\begin{array}{l}
y'^t_1=u^t+b \\
y^t_1=y'^t_1\cdot(1+\sqrt{2D^C_M}\xi^C_{M1})(1+\sqrt{2D^U_M}\xi^U_{M1}) + \\
 \sqrt{2D^C_A} \xi^C_{A1} + \sqrt{2D^U_A} \xi^U_{A1}.
\end{array}
\end{equation}
Then the expected value and variance in the input layer are:
\begin{equation}\label{eq:AppA_y1_mean_var}
\begin{array}{c}
E(y^t_1)=E(y'^t_1)=u^t+b\\
Var(y^t_1) = \sigma^2_{add} + \sigma^2_{mult} E^2(y^t_1),
\end{array}
\end{equation}
where $\sigma^2_{add}=2D^C_A+2D^U_A$, $\sigma^2_{mult}=2D^C_M+2D^U_M+4D^C_M D^U_M$. It is the same as for FNN.

Due to the global coupling the state equation for recurrent layer before the noise influence is the same for each $i$th node:
\begin{equation}
y'^t_2=y'^t_{2,i} = \alpha\big( \gamma y^t_1 + \beta\frac{1}{I_2}\sum^{I_2}_{j} y^{t-1}_{2,j} +b \big).
\end{equation}
The equation after noise influence is:
\begin{equation}
\begin{array}{l}
y^t_{2,i}=y'^t_{2} \cdot(1+\sqrt{2D^C_M}\xi^C_{M2})(1+\sqrt{2D^U_M}\xi^U_{M2,i}) + \\
 \sqrt{2D^C_A} \xi^C_{A2} + \sqrt{2D^U_A} \xi^U_{A2,i}.
\end{array}
\end{equation}
Then the equation for the state without noise can be transformed as:
\begin{equation}\label{eq:AppB_y2}
\begin{array}{l}
y'^t_2=\alpha(\gamma y^t_1 + b) + \alpha\beta\big[ \sqrt{2D^C_A}\xi^C_{A2} +\\
 \frac{1}{I_2}\sum^{I_2}_{j=1} \sqrt{2D^U_A}\xi^U_{A2,j} + y'^{t-1}_2(1+\sqrt{2D^C_M}\xi^C_{M2})\times \\
 \big(1+\frac{1}{I_2}\sum^{I_2}_{j=1} \sqrt{2D^U_M}\xi^U_{M2,j}\big)\big] \approx \alpha (\gamma y^t_1+b) + \\
  \alpha\beta\big[ \sqrt{2D^C_A}\xi^C_{A2} + y'^{t-1}_2(1+\sqrt{2D^C_M}\xi^C_{M2}) \big].
\end{array}
\end{equation}
It is the recurrent formula. The equation contains all the noise from the input node and only correlated noise from the recurrent layer at time ($t-1$). The corresponding expected value and variance are:
\begin{equation}\label{eq:AppB_y2_mean}
E(y'^t_2)=E(y^t_{2,i}) = \alpha \big( \gamma E(y^t_1) + \beta E(y'^{t-1}_2) +b  \big) \\
\end{equation}
\begin{equation}\label{eq:AppB_y2_var}
\begin{array}{c}
Var(y'^t_2)\approx \alpha^2\gamma^2 Var(y^t_1) + \alpha^2\beta^2 \big[ 2D^C_A + \\
2D^C_M E^2(y'^{t-1}_2) + (1+2D^C_M) Var(y'^{t-1}_2)  \big].
\end{array}
\end{equation}
It is the recurrent formula working for $1<t\le T$ until $t=1$. The initial state for t=1 has
\begin{equation}\label{eq:AppB_y2_init_mean_var}
\begin{array}{l}
E(y'^1_2)= \alpha \big( \gamma E(y^1_1) +b  \big) = \alpha\gamma(u^1+b) + \alpha b \\
Var(y'^1_2)=\alpha^2\gamma^2 Var(y^1_1) \ \ \ \ \ \ \ \  \text{for}\ t=1.
\end{array} 
\end{equation}

Finally the output neuron before noise impact has the state equation:
\begin{equation}\label{eq:AppB_y3}
\begin{array}{r}
y'^t_3 = \frac{1}{I_2}\sum^{I_2}_{j=1} y^t_{2,j} + b \approx b + \sqrt{2D^C_A}\xi^C_{A2} +\\
 y'^t_2 (1+\sqrt{2D^C_M} \xi^C_{M2}).
\end{array}
\end{equation}
Its variance is
\begin{equation}\label{eq:AppB_y3_var}
\begin{array}{r}
Var(y'^t_3) = 2D^C_A + 2D^C_M E^2(y'^t_2) + (1+2D^C_M) Var(y'^t_2).
\end{array}
\end{equation}

The output signal $y^t_3$ with noise is described by similar equation to (\ref{eq:AppB_y1}). The expected value and variance of the output signal at time $t$ is:
\begin{equation}\label{eq:AppB_y3_noise_mean}
E(y^t_3) = E(y'^t_3)=E(y'^t_2)+b
\end{equation}
\begin{equation}\label{eq:AppB_y3_noise_var}
\begin{array}{l}
Var(y^t_3) = \sigma^2_{add} + \sigma^2_{mult} E^2(y^t_3) + (1+\sigma^2_{mult}) Var(y'^t_3) \approx \\

\sigma^2_{add} + \sigma^2_{mult} E^2(y^t_3) + (1+\sigma^2_{mult})\times \\
 \big[ 2D^C_A + 2D^C_M E^2(y'^t_2) + (1+2D^C_M) Var(y'^t_2)  \big].
\end{array}
\end{equation}
The term in $[\cdots ]$ brackets shows the noise impact from the recurrent layer in time $t$. It is described by the recurrent formula (\ref{eq:AppB_y2_var}) with the noise impact from all the previous time moments. 

\begin{center} \textbf{Renormalization} \end{center}

The range of the output signal in RNN depends not only on the range of the input signal but also on the input sequence itself. The prediction of output limits can be prepared only for known input signal. The mean output signal is (\ref{eq:AppB_y3_noise_mean}) where the mean of recurrent layer $E(y'^t_2)$ can be found using the recurrent formula (\ref{eq:AppB_y2_mean}) with initial state (\ref{eq:AppB_y2_init_mean_var}). It doesn't reliant on noise intensities, but strongly depend on the input sequence. To keep the same scale one can calculate the limits for the renormalization is:
\begin{equation}
\begin{array}{c}\label{eq:AppB_limits}
y_{min}=min\big( E(y^t_3) \big), \ \ \ \ \ \ t_0<t\le T \\
y_{max}=max\big( E(y^t_3) \big), \ \ \ \ \ \ t_0<t\le T 
\end{array}
\end{equation}
starting from some transient time $t_0$. In numerical simulation and analytics the renormalization is used:
\begin{equation}\label{eq:AppB_renorm}
\big( y^t_3  \big)_R = \frac{y^t_3 - y_{min}}{y_{max}-y_{min}}.
\end{equation} 
The variable $y^t_3$ here is multiplied by $1/(y_{max}-y_{min})$.
Then the analytical formula for the output variance in the case of normalization:
\begin{equation}\label{eq:AppB_renorm_var}
Var \big( y^t_3  \big)_R = Var(y^t_3)/({y_{max}-y_{min}})^2.
\end{equation} 

\nocite{*}

\bibliography{references}

\end{document}